\documentclass[twocolumn]{aastex631}
\usepackage{CJK}
\usepackage{amsfonts}
\usepackage{amsmath}
\usepackage{mathrsfs}
\usepackage{epsfig}
\usepackage{subfigure}

\shorttitle{Origin of the Blazhko effect}
\shortauthors{J.-S. Niu}

\begin{document}
\begin{CJK*}{UTF8}{gbsn}

\title{Pulsation harmonics reveal the origin of the century-old Blazhko effect}

\author[0000-0001-5232-9500]{Jia-Shu Niu (牛家树)}
\correspondingauthor{Jia-Shu Niu}
\email{jsniu@sxu.edu.cn}
\affil{Institute of Theoretical Physics, Shanxi University, Taiyuan 030006, China}
\affil{State Key Laboratory of Quantum Optics Technologies and Devices, Shanxi University, Taiyuan 030006, China}
\affil{Collaborative Innovation Center of Extreme Optics, Shanxi University, Taiyuan 030006, China}

\begin{abstract}
The Blazhko effect---a quasi-periodic modulation of pulsation amplitude and phase observed in all subtypes of RR Lyrae stars---has defied physical explanation for over a century. Here we show that pulsation harmonics of the Blazhko RRab star V783~Cyg, traditionally regarded as passive replicas of the fundamental mode, encode a direct observational signature of the effect's origin. Using Kepler short-cadence photometry, we identify a subset of harmonics---here termed ``disharmonized harmonics''---whose amplitude variations are anti-correlated with those of lower-order harmonics and of the fundamental mode. These disharmonized harmonics map precisely onto the base of the convective envelope, where convection--pulsation interaction is strongest, and their behaviour directly traces the strength of turbulent convection. We interpret the Blazhko modulation as arising from quasi-periodic variation of turbulent convection in the stellar envelope, which regulates both the efficiency of energy transport and the thickness of the $\kappa$-excitation zone. This interpretation naturally accounts for the occurrence of the Blazhko effect across RR Lyrae subtypes and its absence when the convective envelope extends only shallowly into the He~I/H ionization zone, establishing an observationally grounded framework that links harmonic behaviour to the convective dynamics underlying this long-standing puzzle.
\end{abstract}

\section{Introduction}
\label{sec:intro}
RR Lyrae stars are low-mass, horizontal-branch pulsators whose self-excited oscillations are driven primarily by the $\kappa$ mechanism in the He~II ionization zone \citep{Baker1962, Cox1963}, with additional driving from the He~I/H region \citep{Bono1994}.
The convective envelope, which can substantially overlap with these ionization zones, plays a critical role in shaping the observed pulsation properties \citep{Unno1967, Baker1979, Houdek2015}.

Despite decades of study, the Blazhko effect \citep{Blazhko1907, Shapley1916, Kurtz2022}---a quasi-periodic modulation of pulsation amplitude and phase on timescales from days to years \citep{Donev2025}---has remained without a confirmed physical origin.
The effect is common, observed in RRab, RRc, and RRd subtypes with incidence rates reaching 50\% or higher \citep{Benko2014, Jurcsik2014, Smolec2017, Netzel2018, Benko2023, Varma2024RRc, Varma2024RRd}, yet no proposed mechanism has achieved consensus.

Models including magnetic obliquity \citep{Shibahashi2000}, nonradial mode excitation \citep{Cox2013, Bryant2015}, resonant mode coupling \citep{Buchler2011, Kollath2011}, pulsation-driven helium transport \citep{Kovacs2026}, and the magnetoconvective mechanism of Stothers \citep{Stothers2006, Stothers2010, Stothers2011}--whose magnetic-field component faces severe challenges \citep{Smolec2016, Kovacs2016}--among others, have been ruled out, face severe challenges \citep{Kolenberg2009, Smolec2011, Molnar2012a}, or at best address only specific aspects of the phenomenology, such as the 9:2 radial mode resonance that explains period-doubling in \textit{Kepler} data \citep{Kolenberg2010, Kollath2011}.
The central obstacle remains the lack of direct observational tracers that can discriminate among proposed mechanisms.

Here we take a different approach: rather than testing models from the top down, we extract information directly from the observed modulations themselves, using pulsation harmonics as probes of the stellar interior.
Pulsation harmonics---integer multiples of the fundamental pulsation frequency---have traditionally been dismissed as passive replicas and excluded from asteroseismic analysis.
However, recent work has shown that in high-amplitude $\delta$ Scuti stars \citep{Niu2023, Niu2024, Xue2024} and in non-Blazhko RR Lyrae stars \citep{Niu2026_RRab_non}, some harmonics exhibit amplitude and frequency variations that are uncorrelated with---or even anti-correlated to---those of the fundamental mode.
We term these ``disharmonized harmonics'': harmonics whose amplitude modulations deviate systematically from the behaviour expected of a passive replica.
Their existence implies that harmonics carry information beyond that of the parent mode and can serve as observational tracers of specific physical processes within the stellar interior.
Moreover, in non-Blazhko RR Lyrae stars, the morphology of the harmonic spectrum maps onto the outer structure of the star \citep{Niu2026_RRab_non}---a correspondence grounded in the well-established stratification of driving, propagating, and dissipative regions within the stellar envelope.
If this correspondence extends to Blazhko stars, harmonics could provide exactly the missing observational link: a direct diagnostic of whether the Blazhko modulation originates in the $\kappa$-driving region, the convective envelope, or elsewhere.

To pursue this possibility, we analyse V783~Cyg (KIC~5559631), a Blazhko RRab star extensively monitored by the \textit{Kepler} space telescope \citep{Plachy2014, Plachy2021}.
Using short-cadence photometry spanning approximately 90 days (Quarter~6), we measured amplitude and frequency variations for harmonics from the fundamental mode $f_0$ up to $47f_0$, following the methodology of \citet{Niu2026_RRab_non} (see Appendix~\ref{app:methods} for details).

\section{Results and Discussion}
\label{sec:results}
\subsection{Harmonic Spectral Morphology: Mapping the Stellar Interior}
\label{subsec:morphology}
\begin{figure*}[htbp!]
\centering
\includegraphics[width=0.8\textwidth]{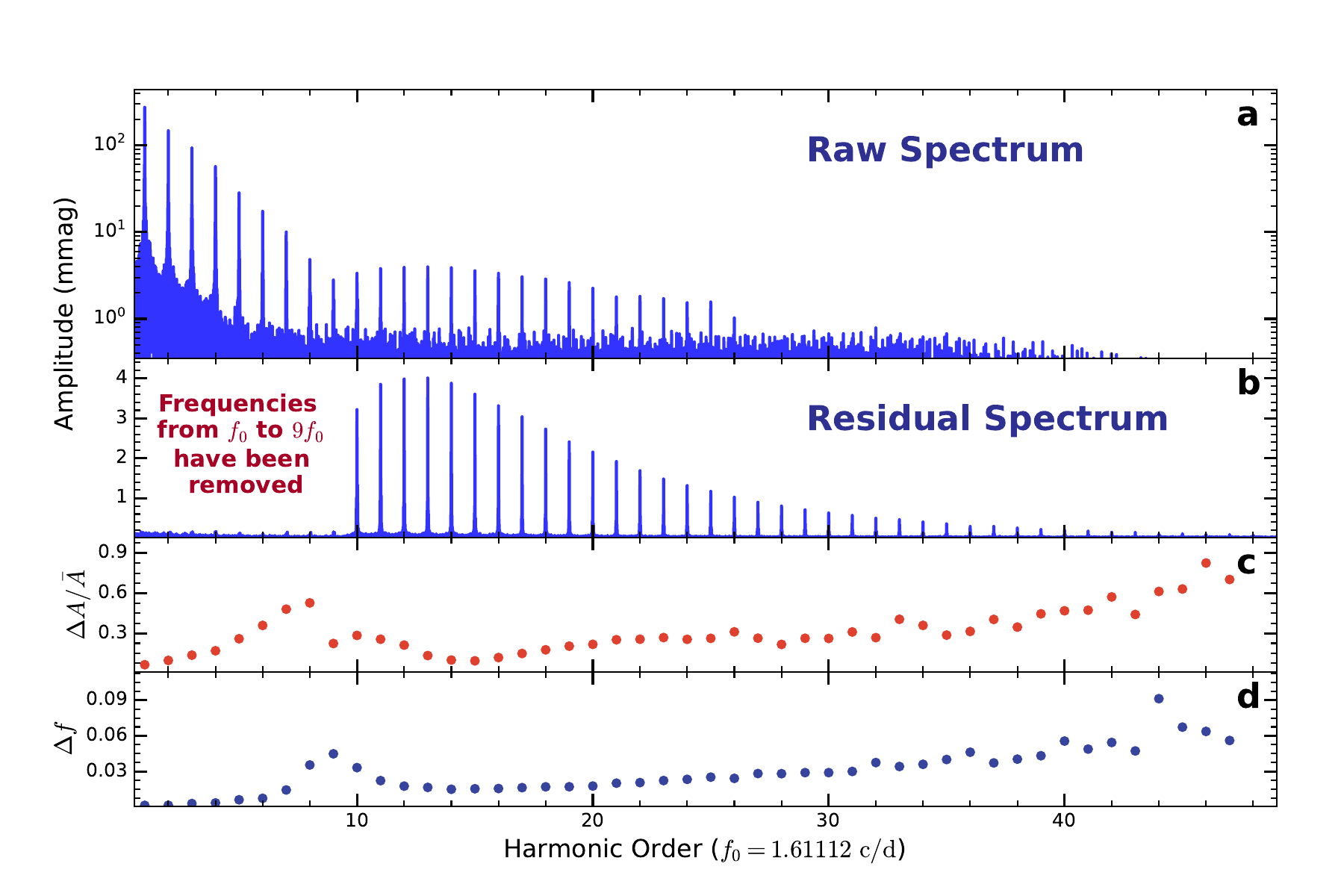}
\caption{Overview of the harmonics $f_0$--$47f_0$ in the frequency spectrum of V783~Cyg and the variability of their amplitude and frequency.
(a) The raw spectrum;
(b) the residual spectrum after pre-whitening the frequencies from $f_0$ to $9f_0$;
(c) the relative amplitude variation ($\Delta A/\bar{A}$) over $\sim 90$ days for each harmonic;
(d) the absolute frequency variation ($\Delta f$) over $\sim 90$ days for each harmonic.
Here, $\Delta A \equiv A_\mathrm{max} - A_\mathrm{min}$ and $\Delta f \equiv f_\mathrm{max} - f_\mathrm{min}$ denote the absolute variations in amplitude and frequency over $\sim 90$ days (see Appendix~\ref{app:figures} Figures~\ref{fig:var_amp_freq01}--\ref{fig:var_amp_freq03}), and $\bar{A}$ denotes the averaged amplitude over the same period.}
\label{fig:spec_res}
\end{figure*}

The frequency spectrum of V783~Cyg (Figure~\ref{fig:spec_res}a, b) reveals a harmonic series extending from the fundamental mode $f_0$ to at least $47f_0$.
The spectral morphology follows a characteristic pattern previously identified in non-Blazhko RRab stars \citep{Benko2016,Niu2026_RRab_non}: an exponential decay across the low-order harmonics ($f_0$--$9f_0$), an asymmetric hump rising near $10f_0$ and peaking around $13f_0$, and a long, slowly decaying tail extending beyond $47f_0$.

This three-segment morphology maps onto the propagation of pulsation waves from the driving region to the stellar surface \citep{Niu2026_RRab_non}: the exponentially decaying low-order harmonics correspond to the driven propagation of the pulsation wave in the radiative zone, where the $\kappa$ mechanism operates; the rising phase of the hump ($10f_0$--$13f_0$) marks the base of the convective envelope, where convection--pulsation interaction injects additional energy into the harmonic spectrum; and the slowly decaying tail (above $15f_0$) traces the propagation of the pulsation wave through the turbulent convective envelope, where the signal is increasingly scattered by convective motions.
This correspondence is grounded in the well-established stratification of driving, propagating, and dissipative regions within the stellar interior: different harmonic orders, by virtue of their spatial resolution within the pulsation cavity, sample different depths, and the spectral morphology reflects the changing balance between driving, convective energy injection, and dissipative scattering as the wave propagates outward.

The amplitude and frequency variations shown in Figure~\ref{fig:spec_res}c, d reveal a further diagnostic: the detailed contributions from different harmonics to the total amplitude and frequency modulations of the light curve.
Compared with non-Blazhko RRab stars \citep{Niu2026_RRab_non}, the most distinct difference in V783~Cyg is the substantially larger $\Delta A/\bar{A}$ values of the exponentially decaying low-order harmonics ($f_0$--$9f_0$).
This enhancement indicates that the $\kappa$-driving zone is significantly modulated during the Blazhko cycle, with the gradual upward trend in $\Delta A/\bar{A}$ across these harmonics pointing specifically to modulation of the He~I/H ionization zone in the outer layers—as interpreted in Section \ref{subsec:variations} below.

\subsection{Disharmonized Harmonics: Direct Tracers of Convection--Pulsation Interaction}
\label{subsec:variations}
\begin{figure*}[htp!]
\centering
\includegraphics[width=0.9\textwidth]{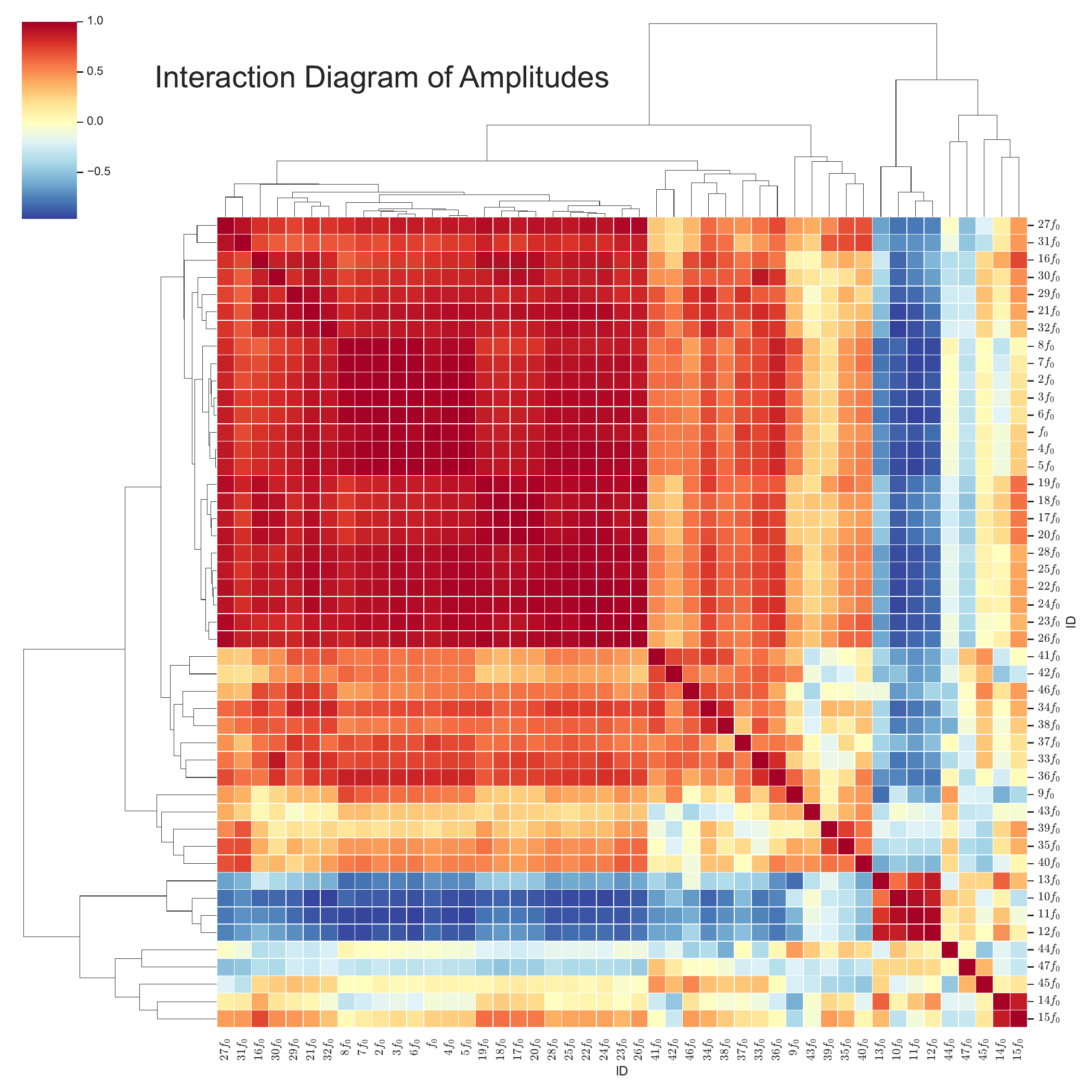}
\caption{Interaction diagram of amplitudes based on $f_0$ and its harmonics up to $47f_0$ in V783~Cyg. Coloured squares represent the correlation coefficient between the amplitudes of the pulsation modes whose columns and rows intersect at that square. The dendrograms on the upper and left sides represent the agglomerative hierarchical clustering process in the amplitude interaction space, grouping harmonics with similar temporal behaviour.}
\label{fig:ID_amp}
\end{figure*}

The temporal amplitude and frequency variations of each harmonic (Appendix~\ref{app:figures}, Figures~\ref{fig:var_amp_freq01}--\ref{fig:var_amp_freq03}) reveal a more nuanced picture than the time-averaged spectrum suggests.
For any given harmonic, both the amplitude and frequency variations are periodic with the Blazhko cycle; however, the amplitude and frequency do not vary in synchrony, and their relative phase offsets differ systematically across harmonic orders.
This behaviour is captured in the interaction diagrams of amplitudes (Figure~\ref{fig:ID_amp}) and frequencies (Appendix~\ref{app:figures}, Figure~\ref{fig:ID_freq}), where harmonics cluster according to the similarity of their temporal evolution\footnote{Details of the interaction diagram are given in Appendix~C of \citet{Niu2022}.}.
The correlation structure of the amplitude variations differs markedly from that of the frequency variations: whereas the frequency variations of nearly all harmonics are mutually correlated (the exceptions, $f_0$ and $2f_0$, being insignificant), the amplitude variations display both correlated and anti-correlated groups.
This difference indicates that the amplitude and frequency variations, while sharing a common origin in the Blazhko modulation, respond to distinct physical mechanisms; because the amplitude variations carry the discriminative information needed to identify those mechanisms, we focus on them hereafter.

\begin{figure*}[htp!]
\centering
\includegraphics[width=0.9\textwidth]{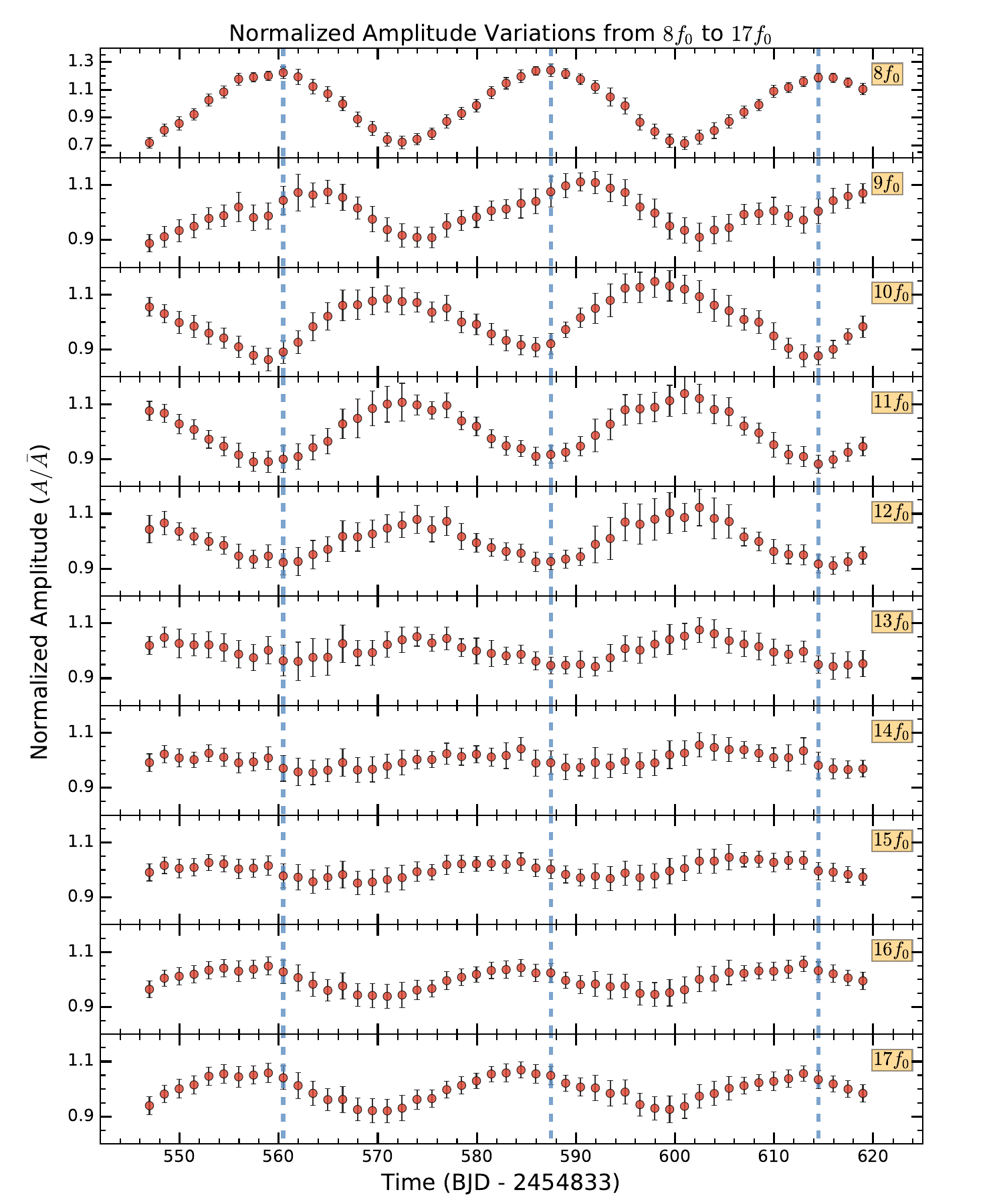}
\caption{Temporal variation of the normalized amplitude ($A/\bar{A}$) for harmonics $8f_0$--$17f_0$. Light blue vertical lines mark the maxima of $A/\bar{A}$ for $8f_0$, serving as baselines for comparing the amplitude variations of different harmonics.}
\label{fig:amp_var_8-17}
\end{figure*}

Figure~\ref{fig:ID_amp} reveals a sharp delineation: harmonics $10f_0$--$13f_0$---precisely those corresponding to the rising hump---are anti-correlated in amplitude with most other harmonics.
These are the disharmonized harmonics.
The detailed amplitude variations of harmonics $8f_0$--$17f_0$ (Figure~\ref{fig:amp_var_8-17}) display this anti-correlation directly, together with two phase reversals as the harmonic order increases.
Because the harmonics $10f_0$--$13f_0$ corresponds to the base of the convective envelope---the region where convection--pulsation interaction is strongest---these disharmonized harmonics serve as direct observational tracers of convective strength.
Because the total amplitude modulation of the Blazhko effect in the light curve is dominated by the low-order harmonics ($f_0$--$9f_0$), the anti-correlation between these two groups of harmonics implies an inverse relationship: as convection intensifies, the pulsation driving is damped.
This inverse relationship between convective strength and pulsation driving efficiency provides the observational foundation for the physical interpretation developed below.

\subsection{Synthesis: Convection Modulation as the Physical Origin}
\label{subsec:synthesis}
The harmonic morphology and their amplitude variation patterns converge on a coherent physical picture.
We interpret the Blazhko modulation as arising from quasi-periodic variation of turbulent convection in the stellar envelope: weakened turbulent convection reduces energy transport in the envelope, enabling more efficient $\kappa$-excitation and hence enhanced pulsation; strengthened turbulent convection transports energy more efficiently, reducing the energy available for $\kappa$-driving and hence weakening pulsation.\footnote{A correlation between the Blazhko effect and the strength of turbulent convection has also been proposed by \citet{Stothers2006,Stothers2010,Stothers2011} and \citet{Li2022}. Most recently, \citet{Benko2026} confirmed a strong positive correlation between the cycle-to-cycle variation strength and the amplitude of the frequency-modulation component of the Blazhko effect, which has been identified as a tracer of turbulent convection in the stellar envelope \citep{Niu2026_RRab_non}.}

The inverse coupling between convective strength and $\kappa$-driving efficiency operates through two channels: (i) weakened turbulent convection reduces the efficiency of energy transport in the envelope, causing energy to accumulate in the radiative zone and thereby increasing the energy available for $\kappa$-driving; (ii) weakened turbulent convection is accompanied by reduced convective overshooting and a thinning of the convective envelope, which, when the convective envelope extends into the driving region, leads to a thicker $\kappa$-excitation zone at the base of the convective envelope and hence more efficient driving.

Within this picture, the gradual upward trends of $\Delta A/\bar{A}$ (Figure~\ref{fig:spec_res}c) and $\Delta f$ (Figure~\ref{fig:spec_res}d) across the exponentially decaying harmonics ($f_0$--$9f_0$) reflect the increasing influence of convective modulation on $\kappa$-driving as the base of the convective envelope is approached.
The positive correlation between the amplitudes of the low-order harmonics and those in the slowly decaying tail is likewise a natural consequence: weakened turbulent convection reduces not only convective energy transport but also the damping of pulsation propagation through the convective envelope, so that both groups of harmonics are enhanced in phase.
This causal chain is reflected in the phase lag between $8f_0$ and $17f_0$ (Figure~\ref{fig:amp_var_8-17}): $8f_0$ samples the state of the $\kappa$-driving zone, whereas $17f_0$ samples the convective envelope, and the lag reflects the response of the the driving region to the changes in convective envelope.

The convective modulation itself is driven by a self-regulating cycle: a local temperature increase enhances opacity through He~I/H ionization, steepening the temperature gradient and strengthening convection; the strengthened convection then transports energy more efficiently, cooling the envelope, reducing the opacity and temperature gradient, and ultimately weakening the convection itself.
In this cycle, the convective envelope acts as a valve that modulates energy transport efficiency in the outer layers, and its natural period---the thermal relaxation timescale of the convective envelope---falls in the range of tens to hundreds of days \citep{Kupka2017, Kovacs2025}, matching the observed timescales of the Blazhko effect. 
The quasi-periodic rather than strictly periodic nature of the modulation is consistent with the stochastic component of turbulent convection, which perturbs the cycle period from one iteration to the next.

The onset of convective envelopes in stars frequently coincides with a sharp rise in local opacity, which typically occurs in the partial ionization regions of helium and hydrogen. Consequently, convective modulation is in principle expected to be ubiquitous across a wide variety of pulsating stars, and its signatures may be present in their pulsation properties to varying degrees.
Indeed, in a large fraction of non-Blazhko RRab stars observed by \textit{Kepler}, the harmonics around the onset and in the tail of the hump structure show significant amplitude and frequency variations \citep{Niu2026_RRab_non}---consistent with convective modulation operating in their envelopes but at a level insufficient to produce the significant Blazhko phenomenology.
A key prediction of this framework is that the distinction between Blazhko and non-Blazhko RR Lyrae stars lies in the spatial relationship between the convective envelope and the pulsation driving region: the Blazhko effect arises when the convective envelope extends deeply into the He~I/H ionization zone (as in V783~Cyg), so that convective modulation directly modulates $\kappa$-driving efficiency.
Conversely, when the convective envelope extends only shallowly into the driving region (as in the non-Blazhko RRab stars of \citet{Niu2026_RRab_non}), convective modulation affects primarily the propagation of pulsation waves through the envelope and only weakly modulates $\kappa$-driving efficiency, so that the Blazhko effect remains insignificant or unobservable.

\section{Conclusion}
\label{sec:con}
The Blazhko effect has resisted explanation for over a century, primarily because no direct observational tracer has been available to discriminate among proposed mechanisms.
We have shown that such a tracer exists in the pulsation harmonics of the Blazhko RRab star V783~Cyg: a subset we term ``disharmonized harmonics''---those whose amplitude variations are anti-correlated with the fundamental mode---map onto the base of the convective envelope, where convection--pulsation interaction is strongest, and directly trace the strength of turbulent convection.

We interpret the Blazhko modulation as arising from quasi-periodic variation of turbulent convection in the stellar envelope, which regulates both the efficiency of energy transport and the thickness of the $\kappa$-excitation zone.
This interpretation accounts for the occurrence of the Blazhko effect across RR Lyrae subtypes and predicts that its strength depends on the depth of overlap between the convective envelope and the He~I/H ionization zone, so that convective modulation directly modulates $\kappa$-driving efficiency.
The present results establish the observational basis for identifying convective modulation as the physical origin of the Blazhko effect; however, confirmation across a broader sample and quantitative theoretical modelling are needed to determine the generality of this mechanism and to reproduce the detailed harmonic variations reported here.

\section*{Acknowledgments}
J.S.N. is grateful to Jue-Ran Niu for a supportive and productive working environment.
The author acknowledges the \textit{Kepler} Science Team and all individuals who contributed to the success of the \textit{Kepler} mission.
All \textit{Kepler} data used in this paper are publicly available through the Mikulski Archive for Space Telescopes (MAST) at \dataset[10.17909/T9059R]{http://dx.doi.org/10.17909/T9059R}.

\software{{\tt astropy} \citep{Astropy2013,Astropy2018,Astropy2022}, {\tt Lightkurve} \citep{lightkurve}, {\tt NumPy} \citep{numpy}, {\tt SciPy} \citep{scipy}, {\tt matplotlib} \citep{matplotlib}, {\tt seaborn} \citep{seaborn}}

\clearpage

\appendix
\setcounter{figure}{0}
\setcounter{table}{0}
\renewcommand{\thefigure}{A\arabic{figure}}
\renewcommand{\thetable}{A\arabic{table}}

\section{Methods}
\label{app:methods}
\subsection{Photometric Data Reduction}
\label{app:methods01}
\begin{figure*}[htp]
\centering
\includegraphics[width=0.8\textwidth]{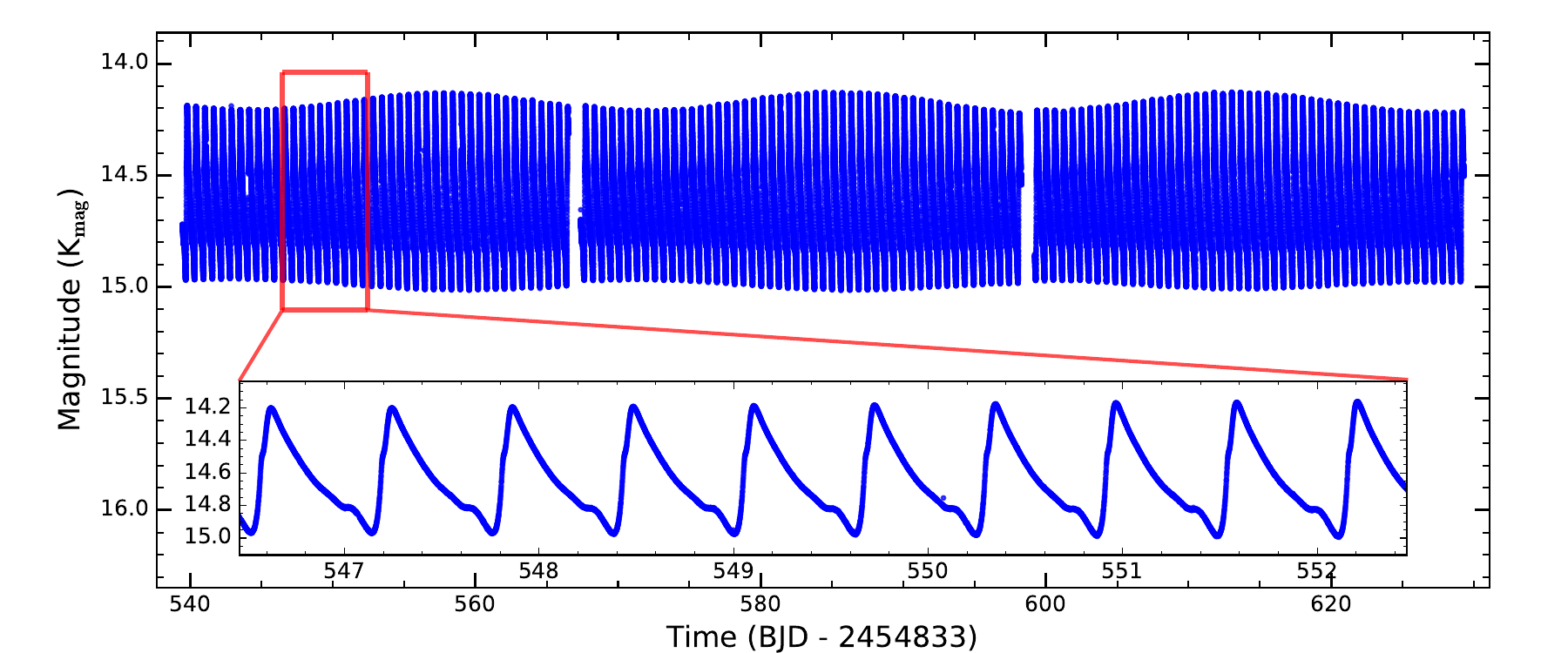}
\caption{Light curves of V783~Cyg based on SC data from Quarter~6. The region within the red rectangle is magnified to show finer details of the light curve structure.}
\label{fig:overview}
\end{figure*}

This work uses short-cadence (SC) photometric data of V783~Cyg from the \textit{Kepler} space telescope, covering BJD~2455372--2455462 (Quarter~6). We used the publicly available Pre-search Data Conditioning (PDC) light curves \citep{Kepler01, Kepler02}, retrieved from the Mikulski Archive for Space Telescopes (MAST)\footnote{\url{http://archive.stsci.edu/kepler}}.
The light curves are provided in barycentric Julian date (BJD) and \textit{Kepler} magnitudes; for each sub-quarter, we normalized the flux and converted it to the \textit{Kepler} magnitude system. Figure~\ref{fig:overview} presents an overview of the SC photometric data.
Quarter~6 was selected because it provides the longest contiguous segment of short-cadence data for V783~Cyg with minimal instrumental gaps, covering approximately 3.25 Blazhko cycles (the Blazhko period of V783~Cyg is $\sim$27.67 days \citep{Plachy2014}).

\begin{figure*}[htp]
  \centering
  \includegraphics[width=0.48\textwidth]{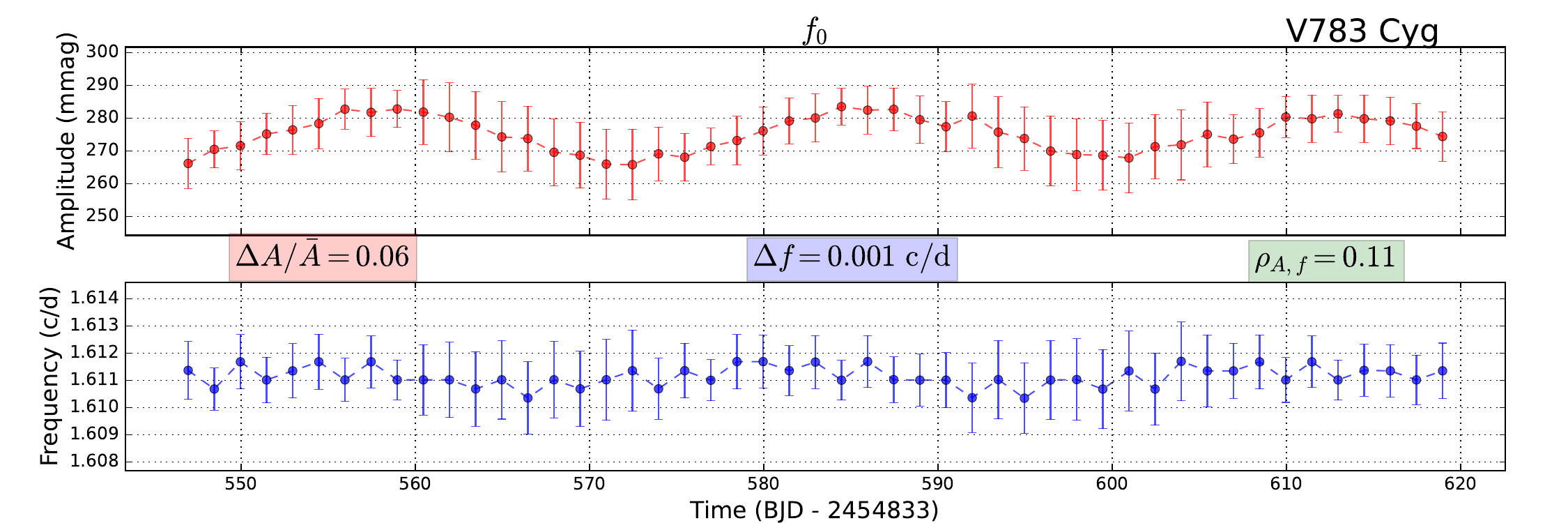}
  \includegraphics[width=0.48\textwidth]{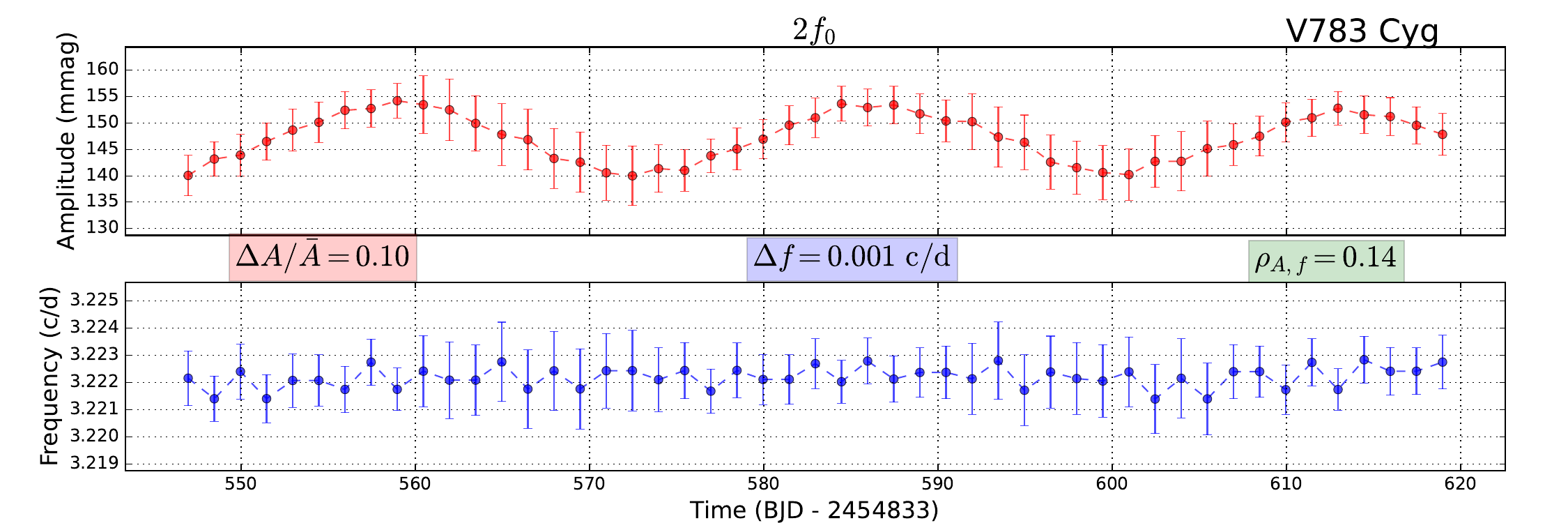}
  \includegraphics[width=0.48\textwidth]{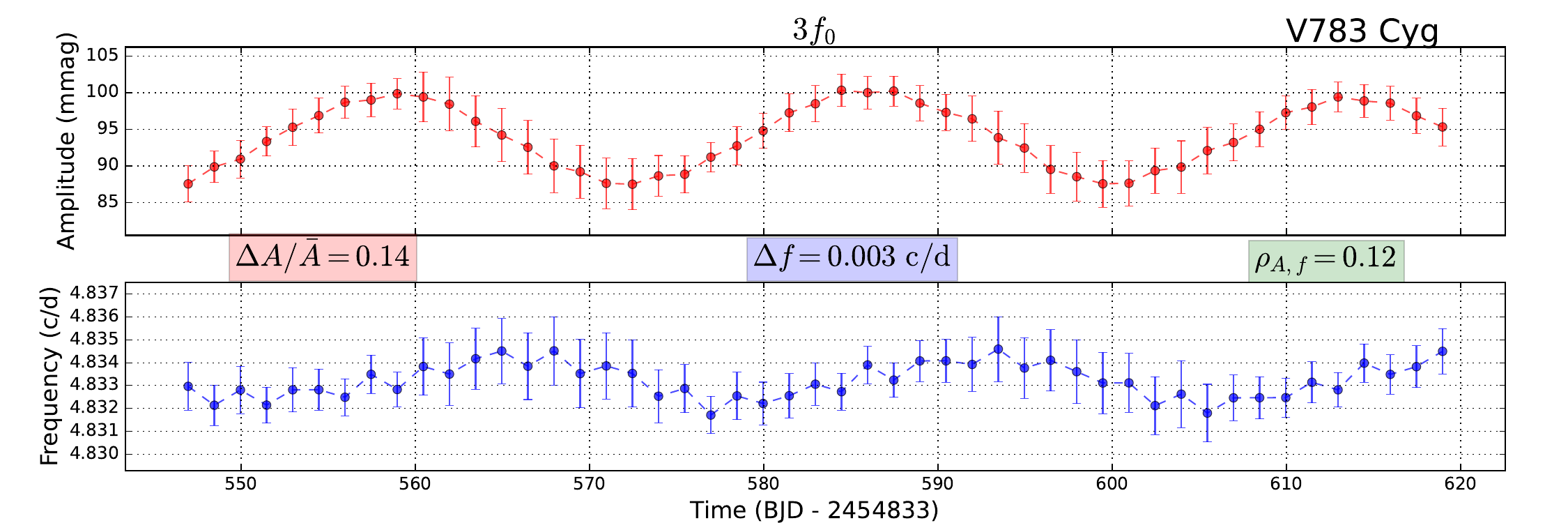}
  \includegraphics[width=0.48\textwidth]{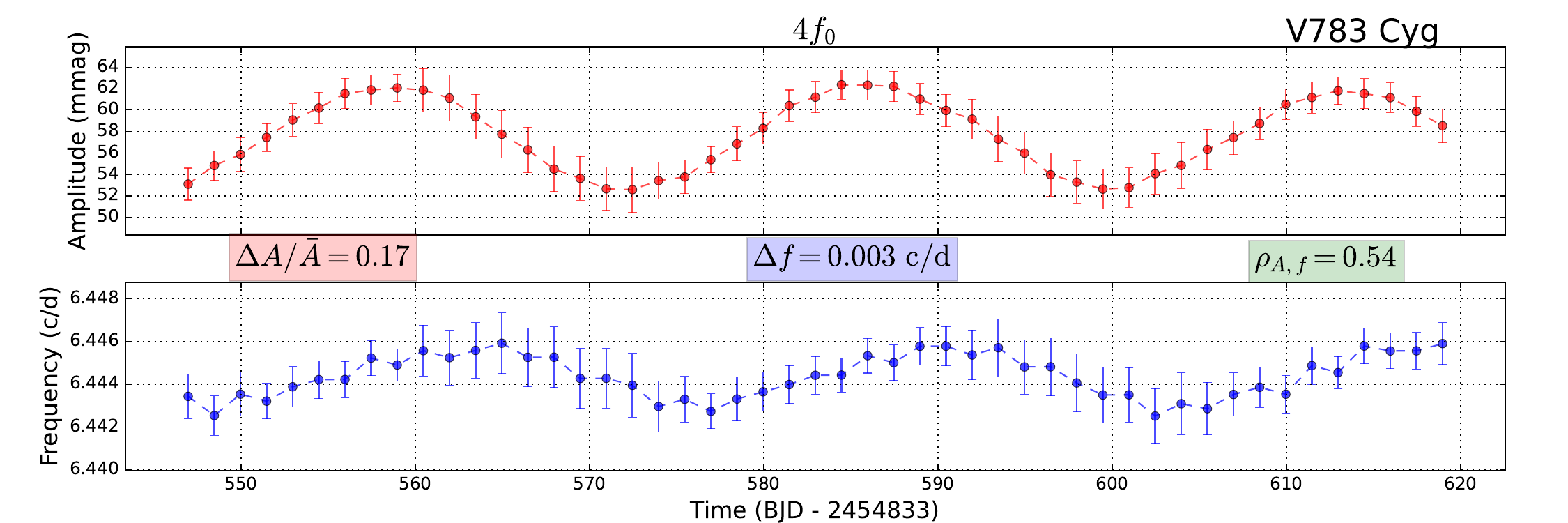}
  \includegraphics[width=0.48\textwidth]{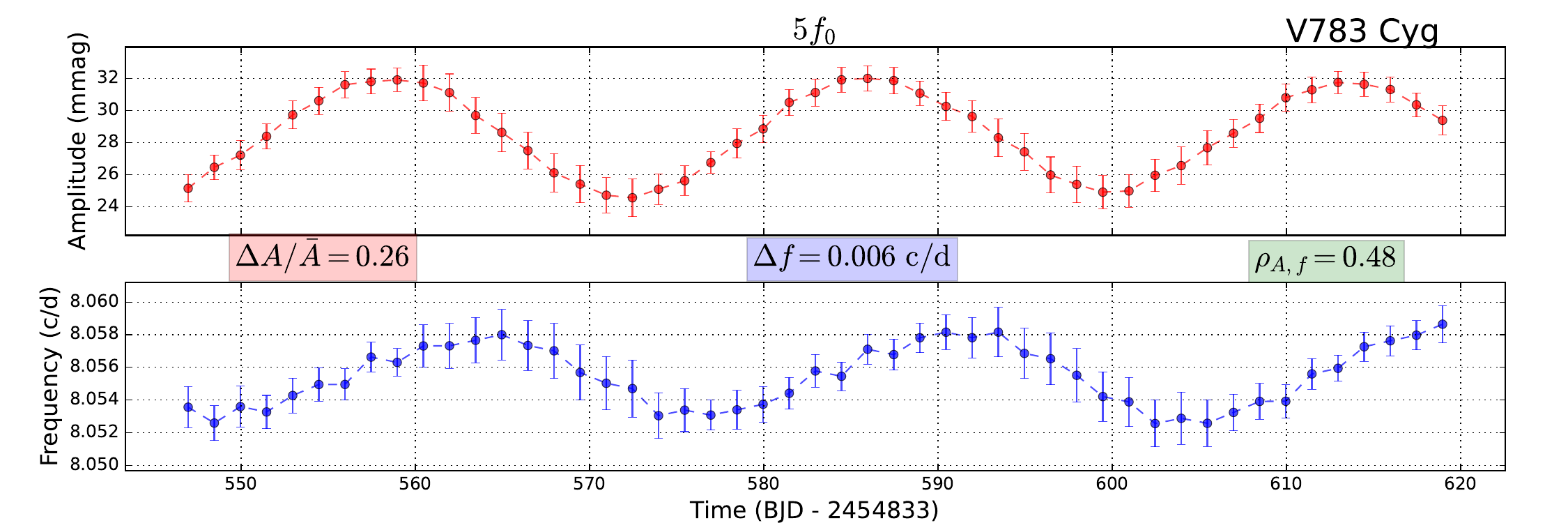}
  \includegraphics[width=0.48\textwidth]{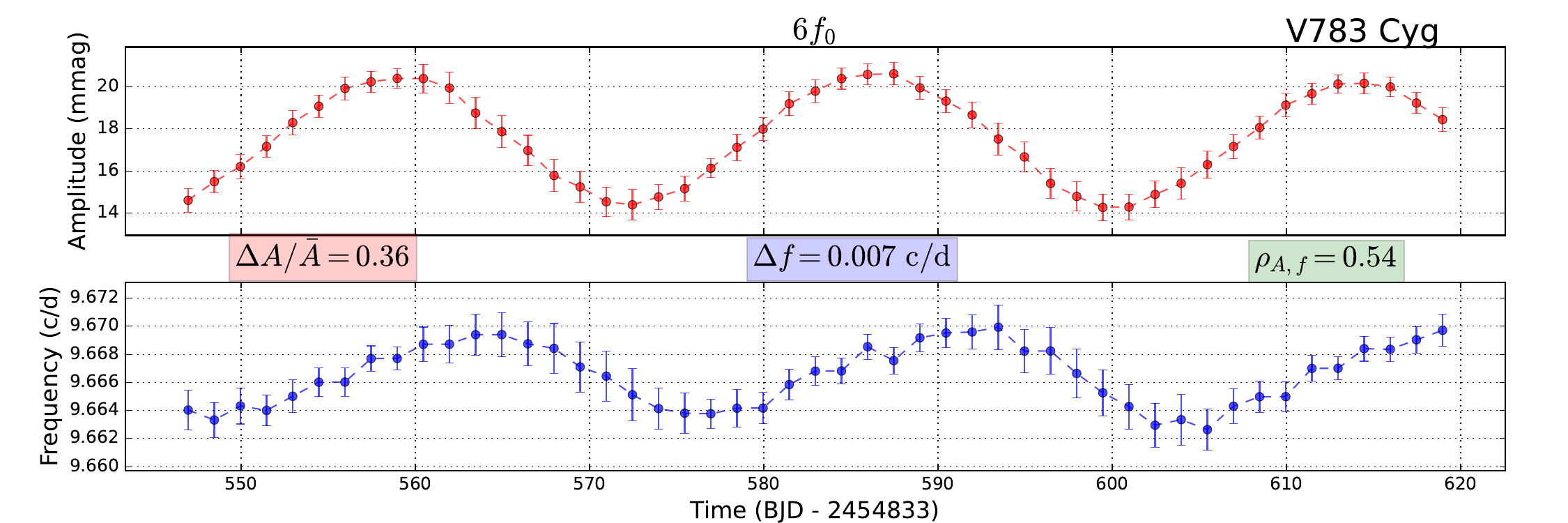}
  \includegraphics[width=0.48\textwidth]{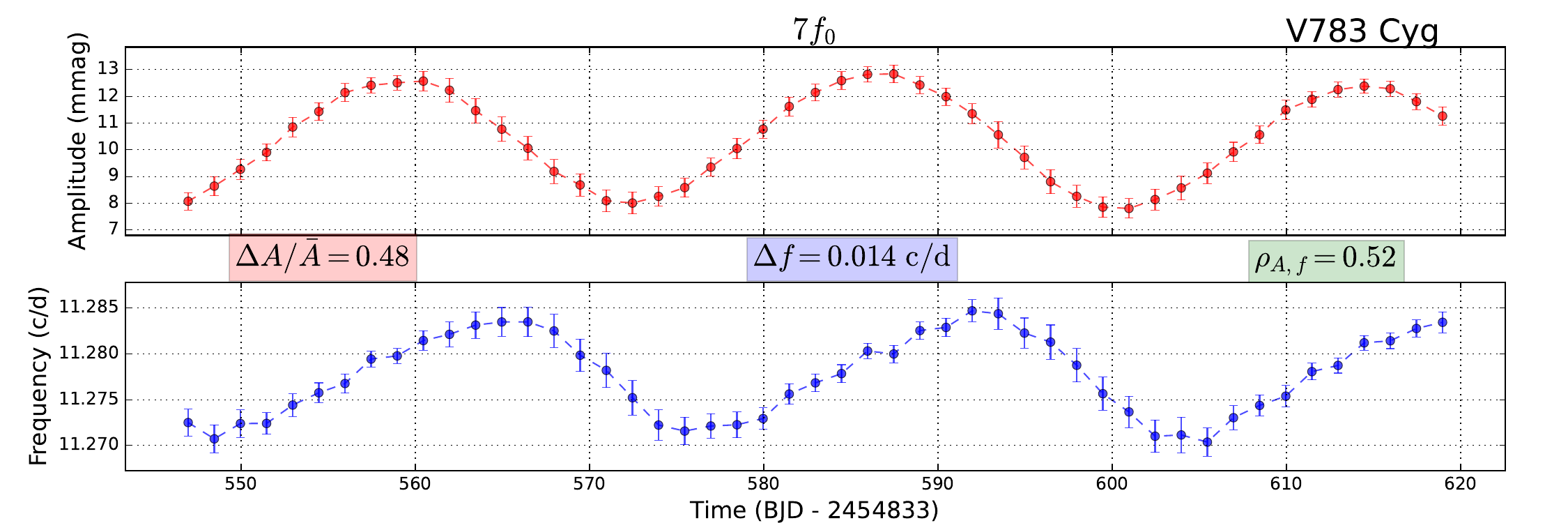}
  \includegraphics[width=0.48\textwidth]{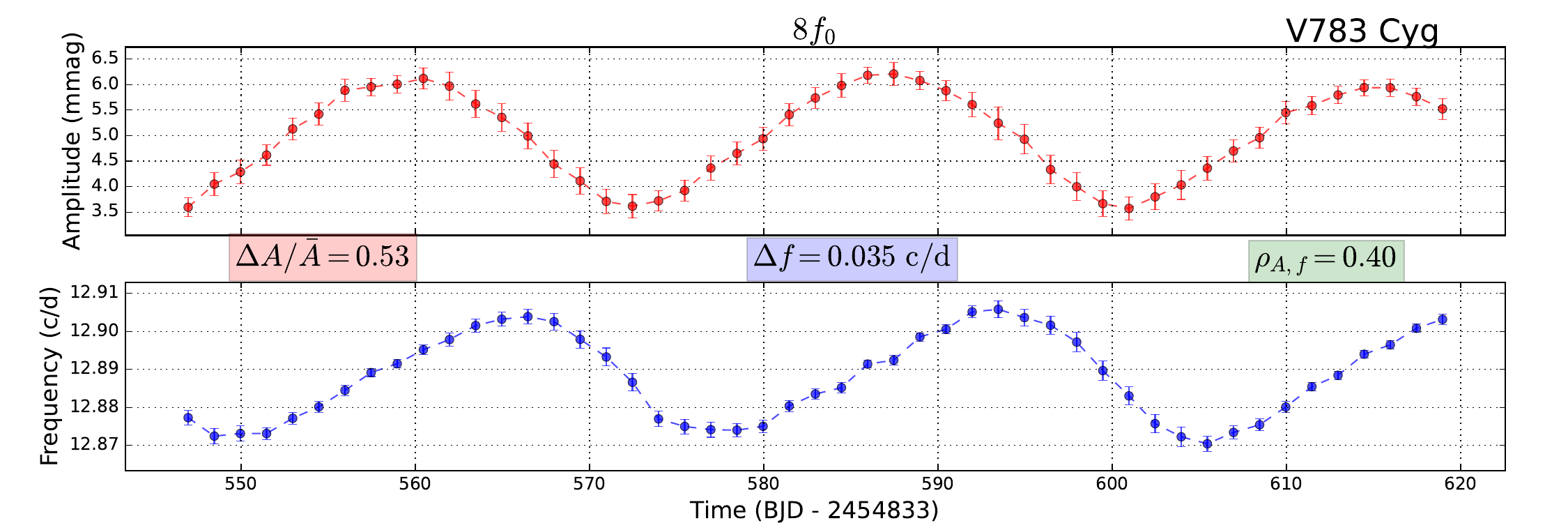}
  \includegraphics[width=0.48\textwidth]{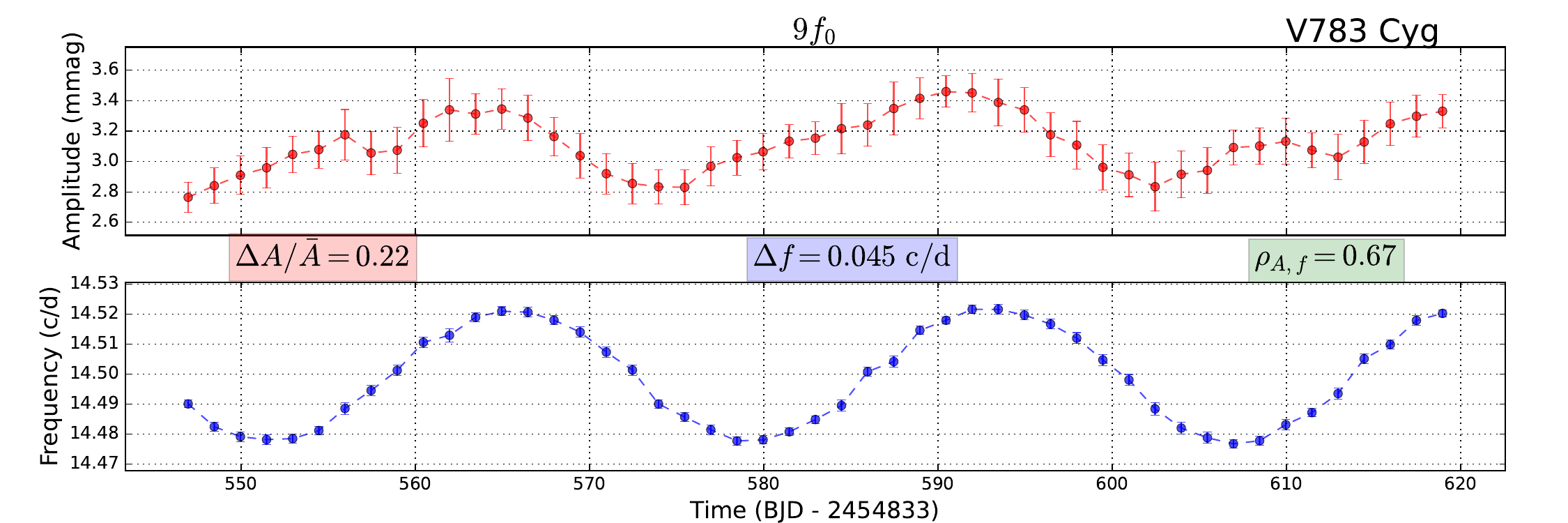}
  \includegraphics[width=0.48\textwidth]{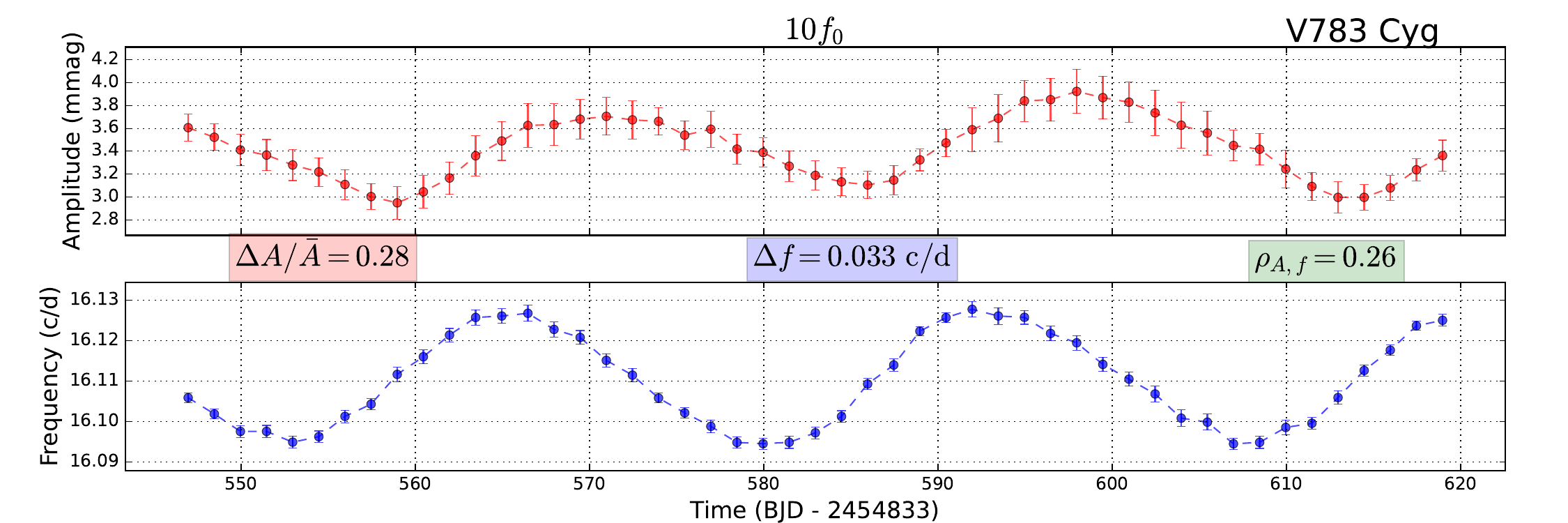}
  \includegraphics[width=0.48\textwidth]{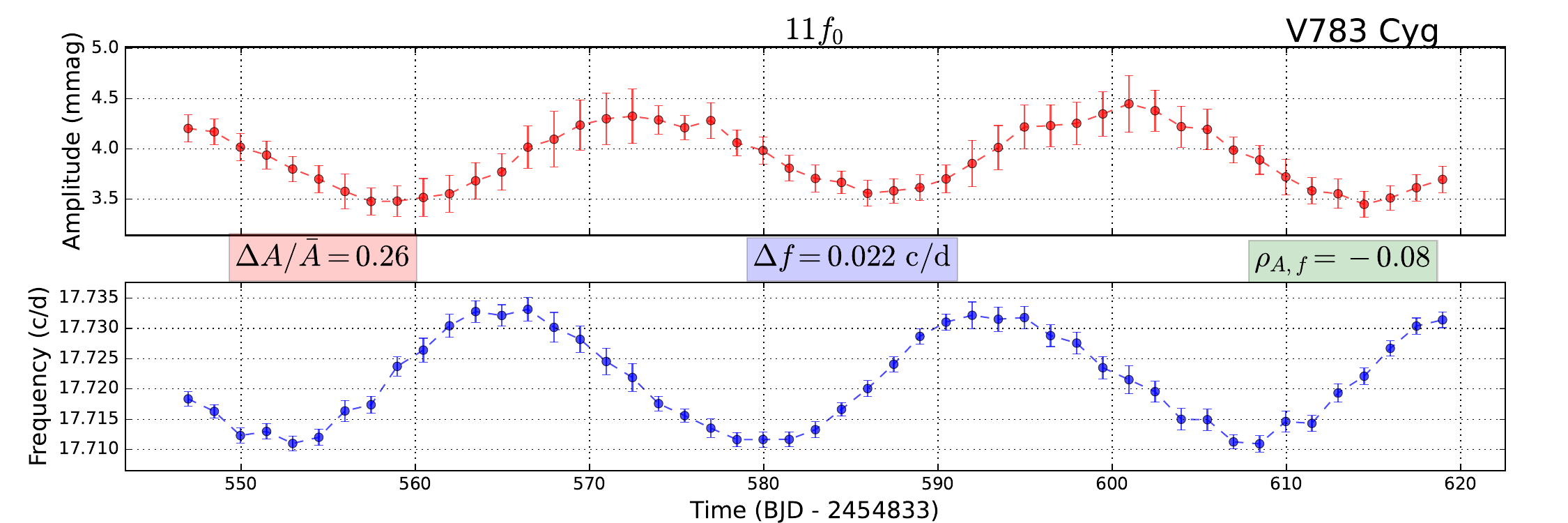}
  \includegraphics[width=0.48\textwidth]{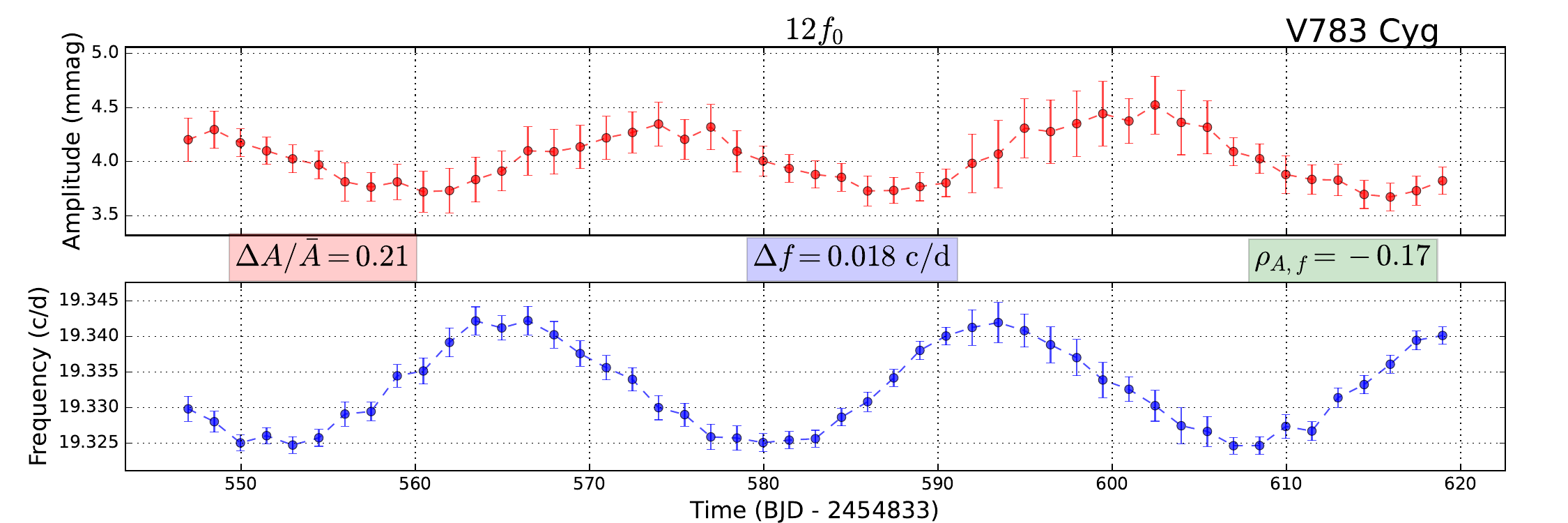}
  \includegraphics[width=0.48\textwidth]{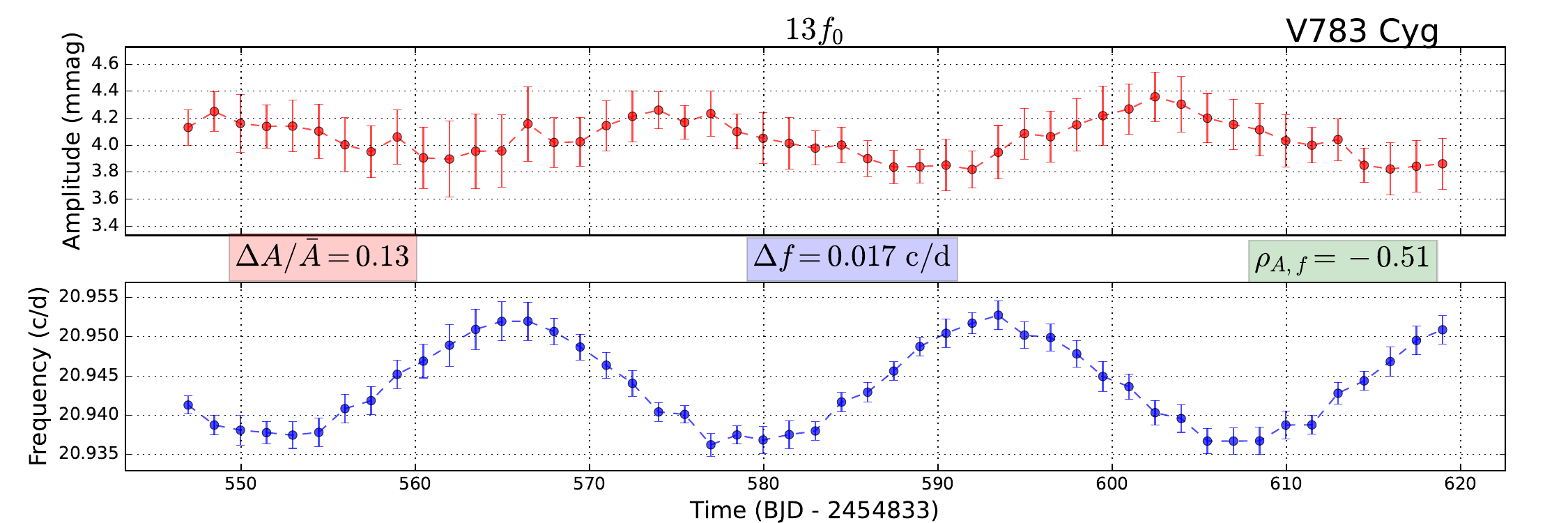}
  \includegraphics[width=0.48\textwidth]{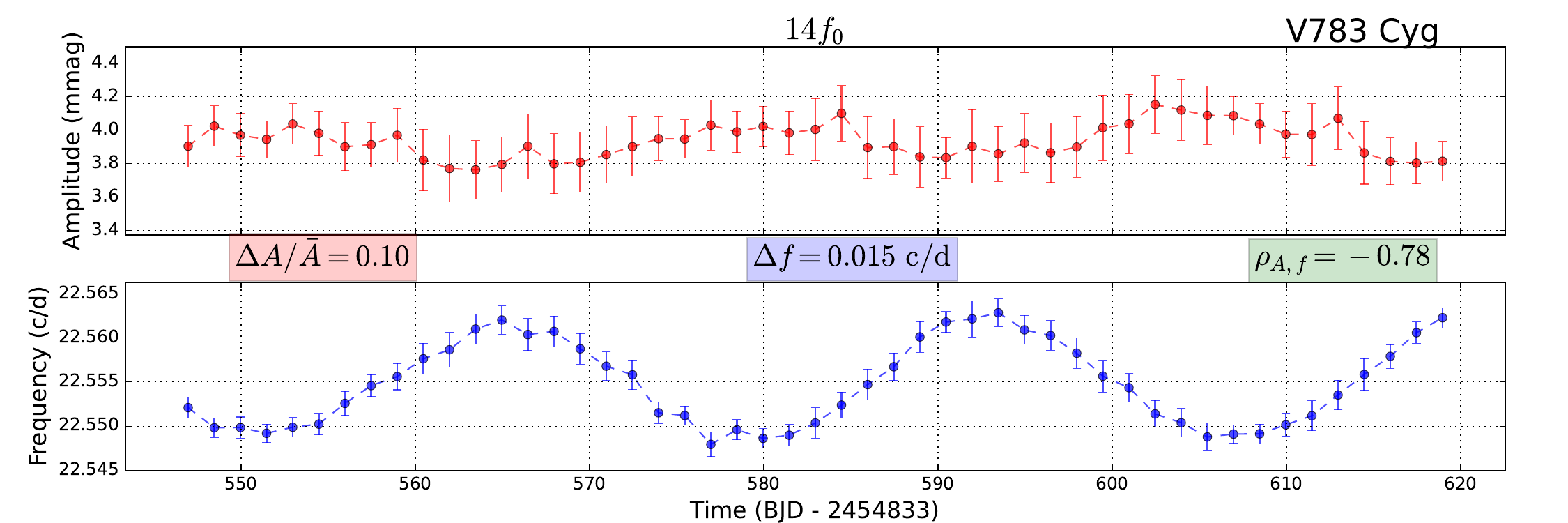}
  \caption{Temporal variations in amplitude and frequency for harmonics $f_0$--$47f_0$, part I.}
  \label{fig:var_amp_freq01}
\end{figure*}

\begin{figure*}[htp]
  \centering
  \includegraphics[width=0.48\textwidth]{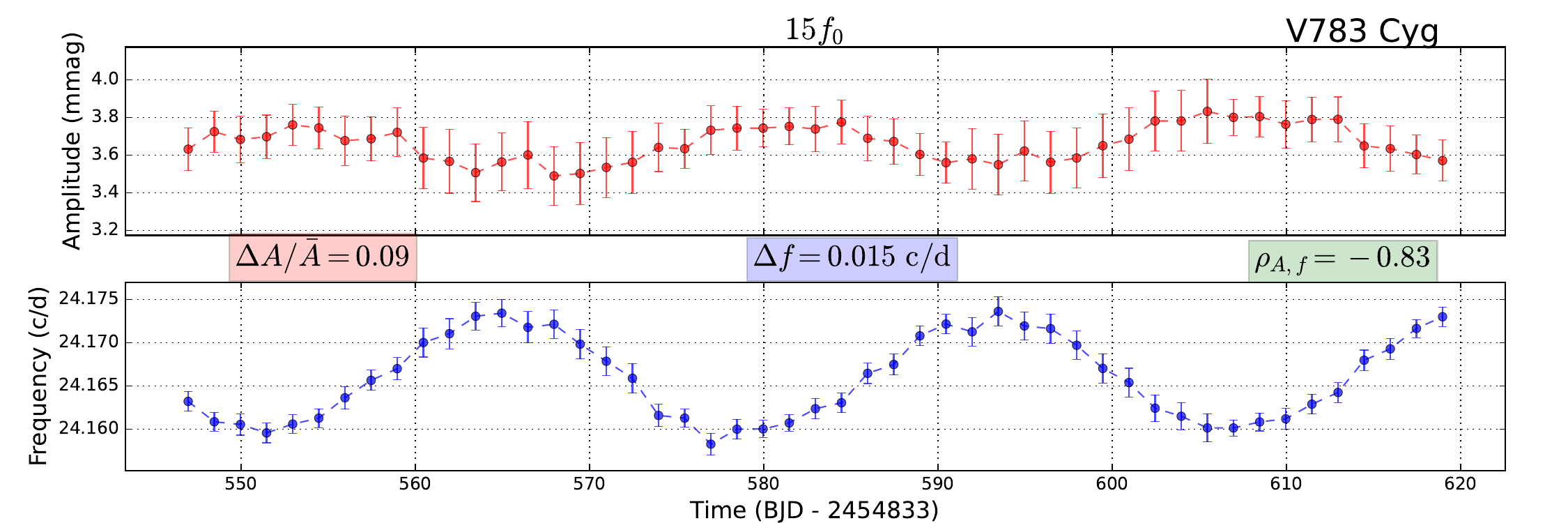}
  \includegraphics[width=0.48\textwidth]{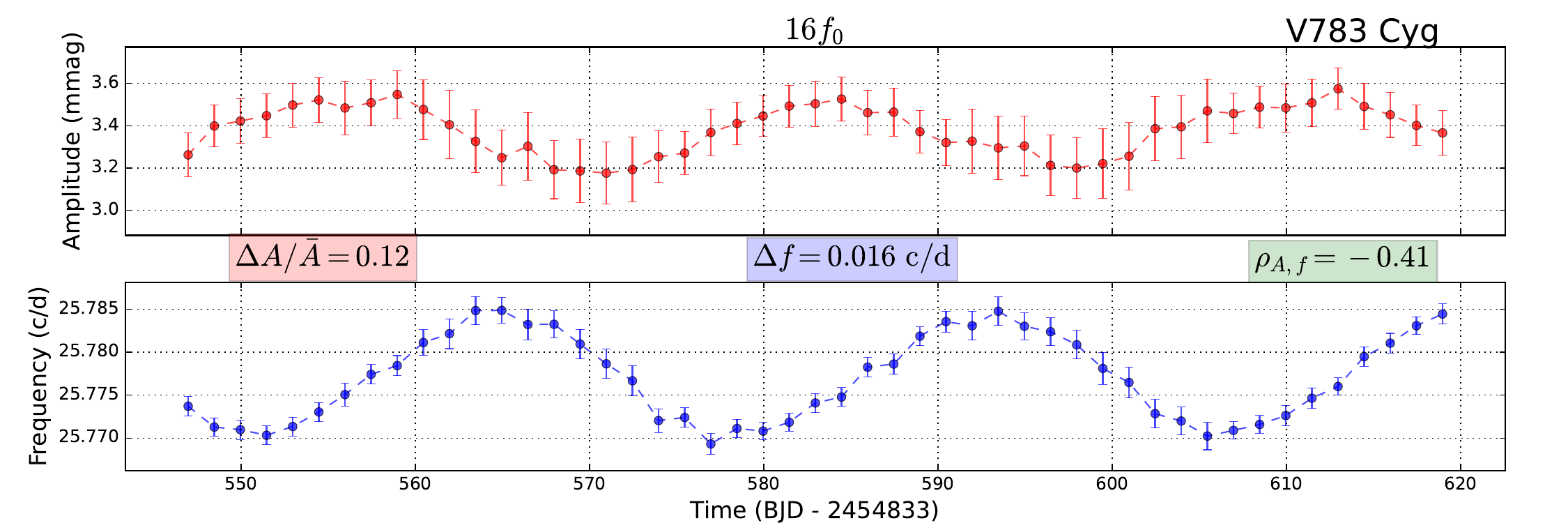}
  \includegraphics[width=0.48\textwidth]{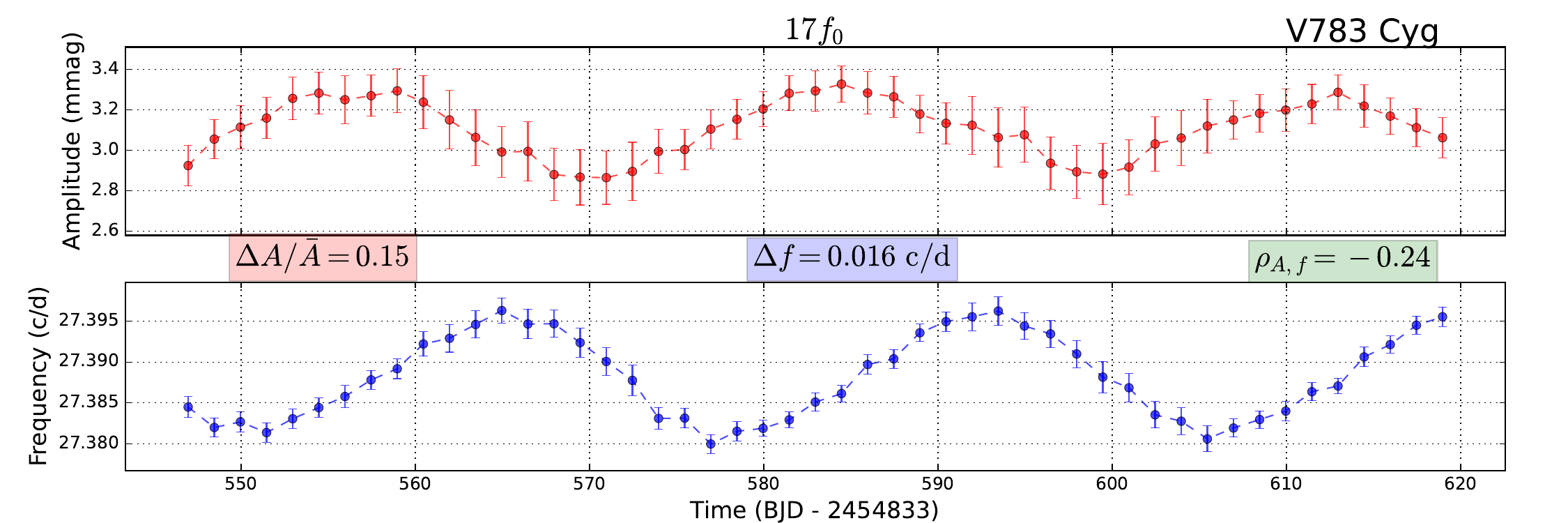}
  \includegraphics[width=0.48\textwidth]{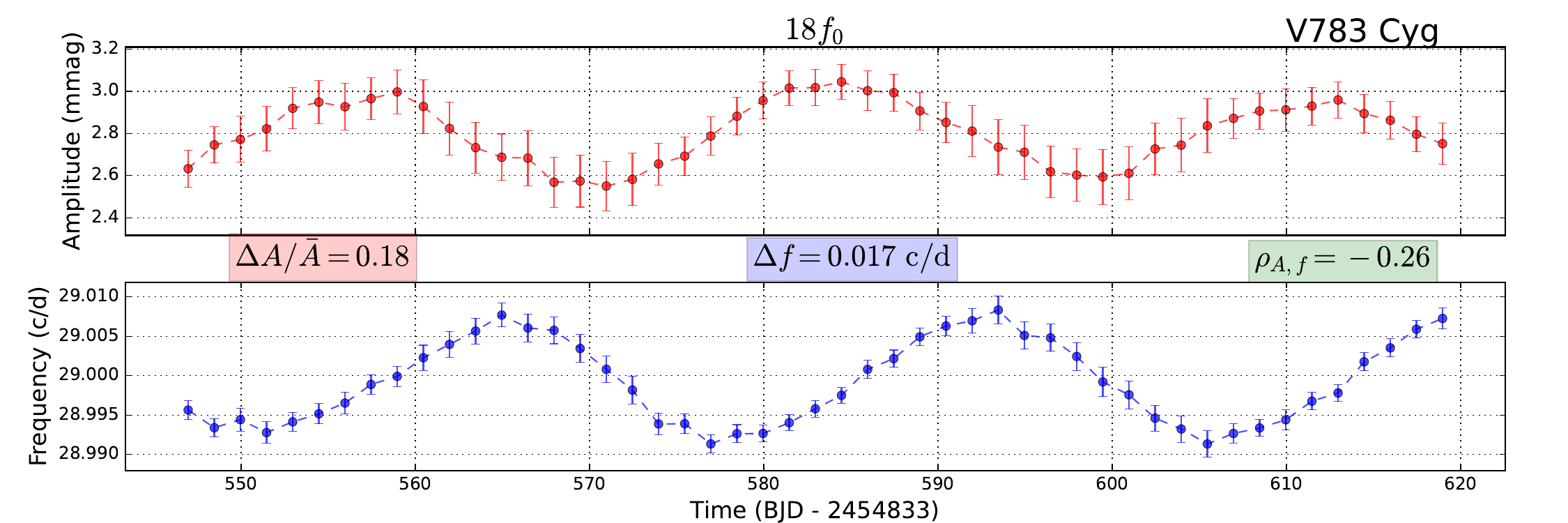}
  \includegraphics[width=0.48\textwidth]{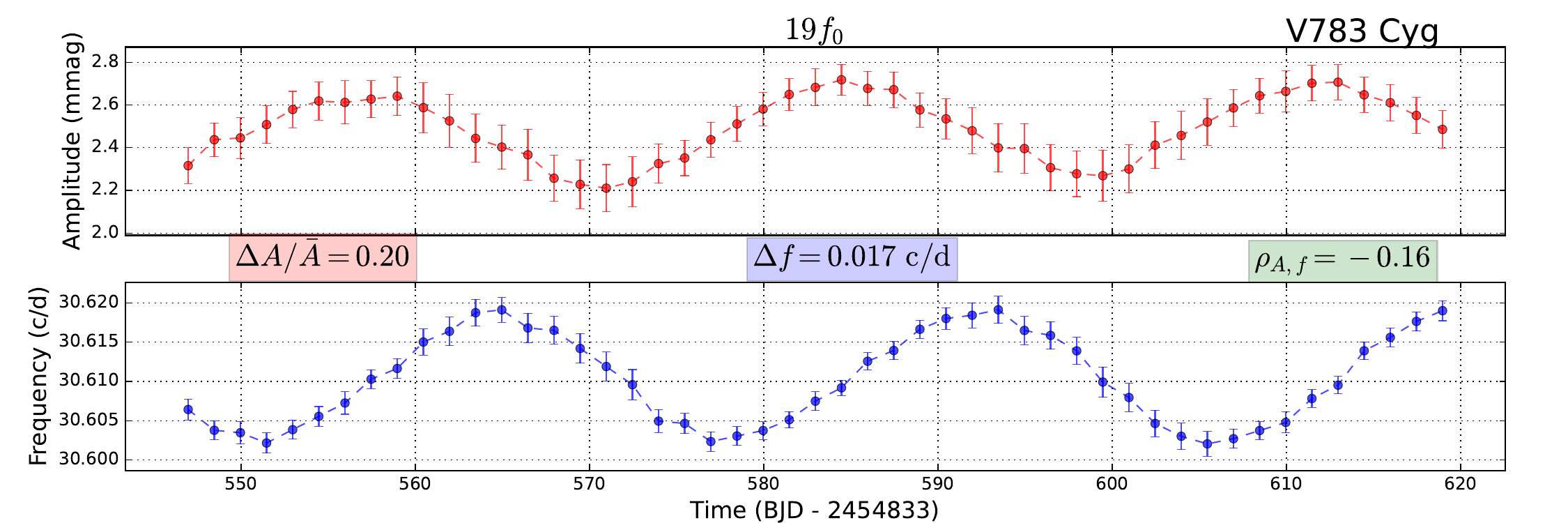}
  \includegraphics[width=0.48\textwidth]{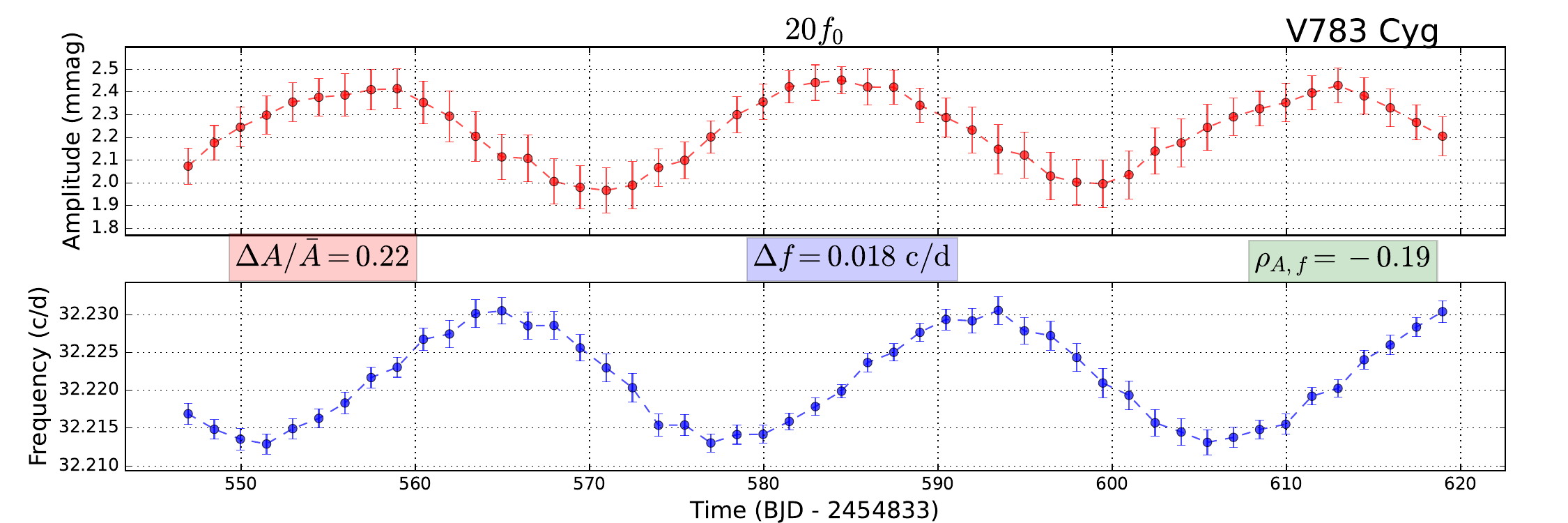}
  \includegraphics[width=0.48\textwidth]{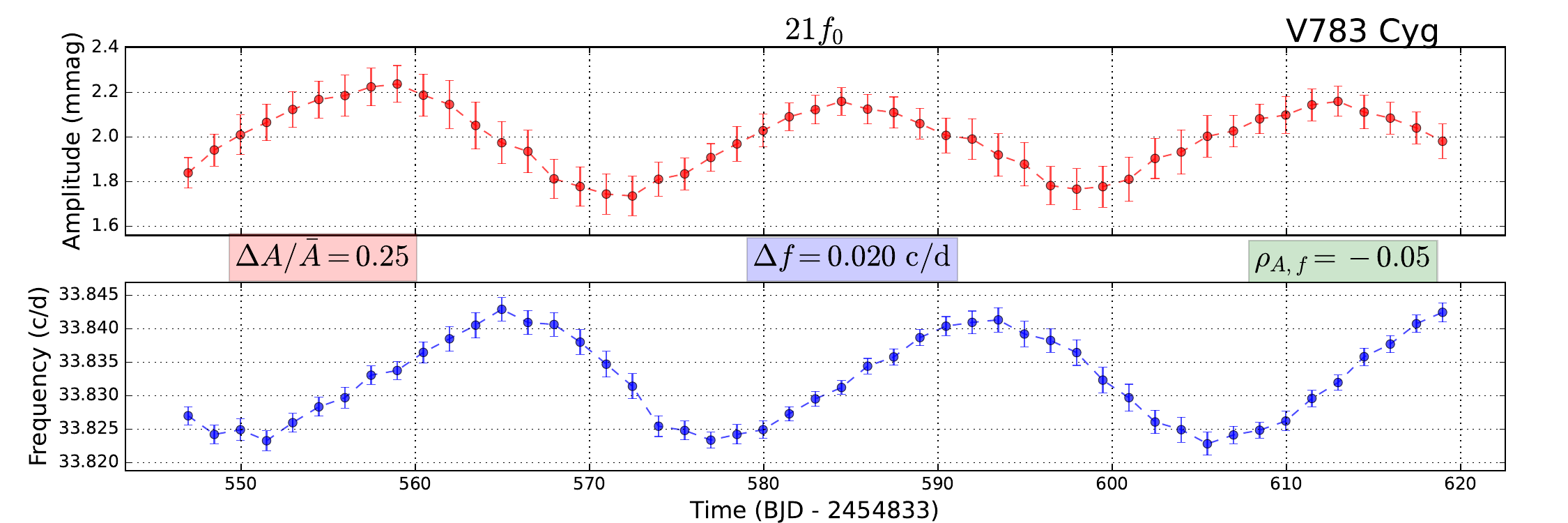}
  \includegraphics[width=0.48\textwidth]{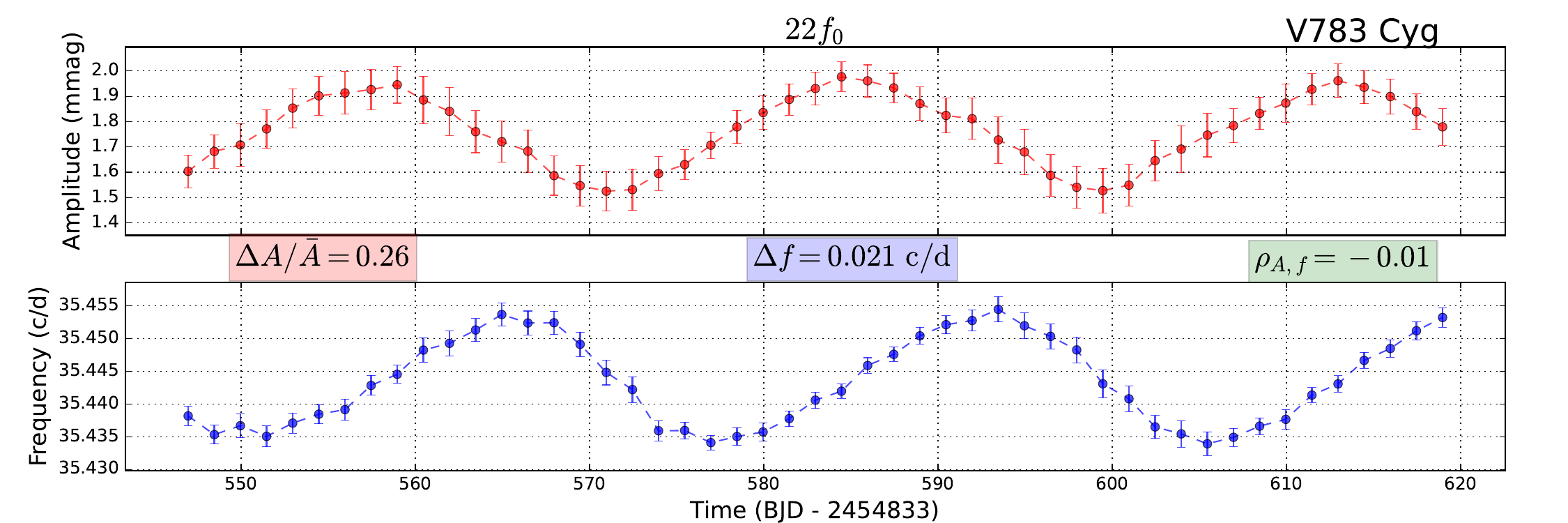}
  \includegraphics[width=0.48\textwidth]{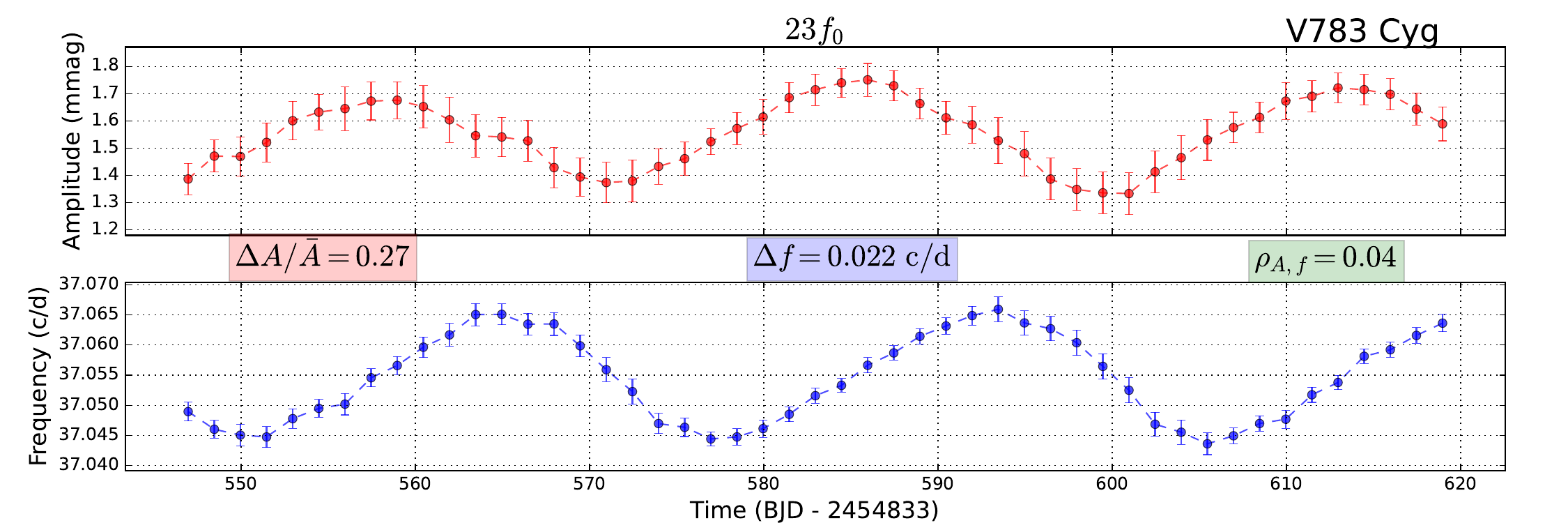}
  \includegraphics[width=0.48\textwidth]{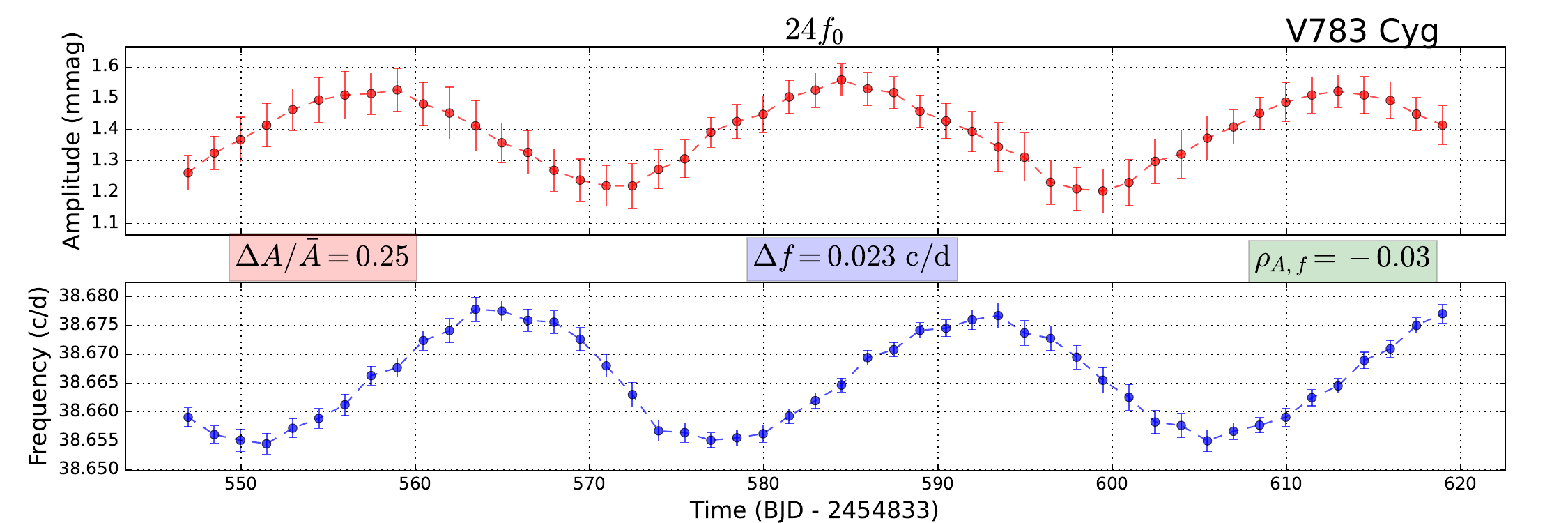}
  \includegraphics[width=0.48\textwidth]{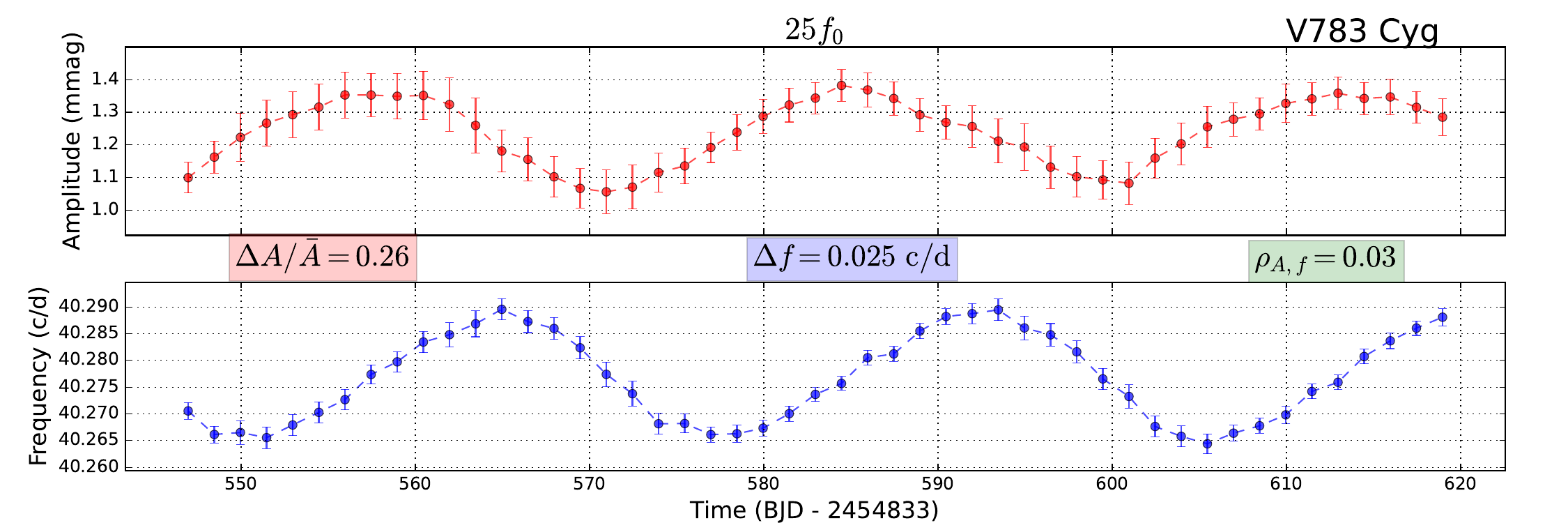}
  \includegraphics[width=0.48\textwidth]{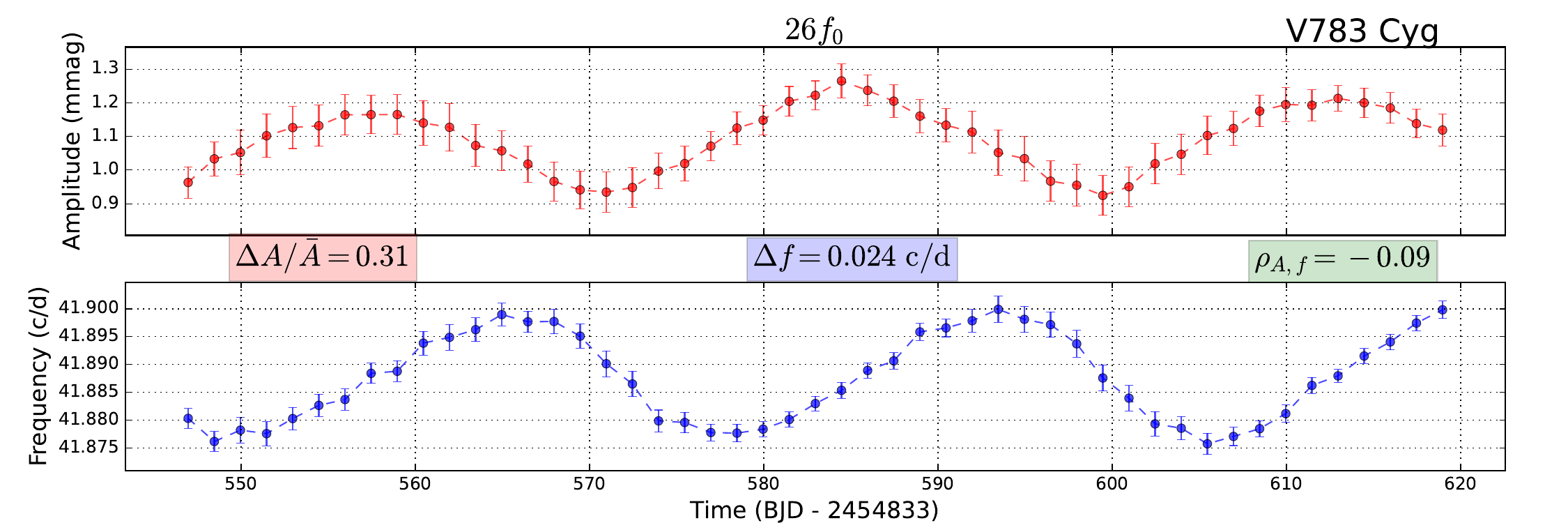}
  \includegraphics[width=0.48\textwidth]{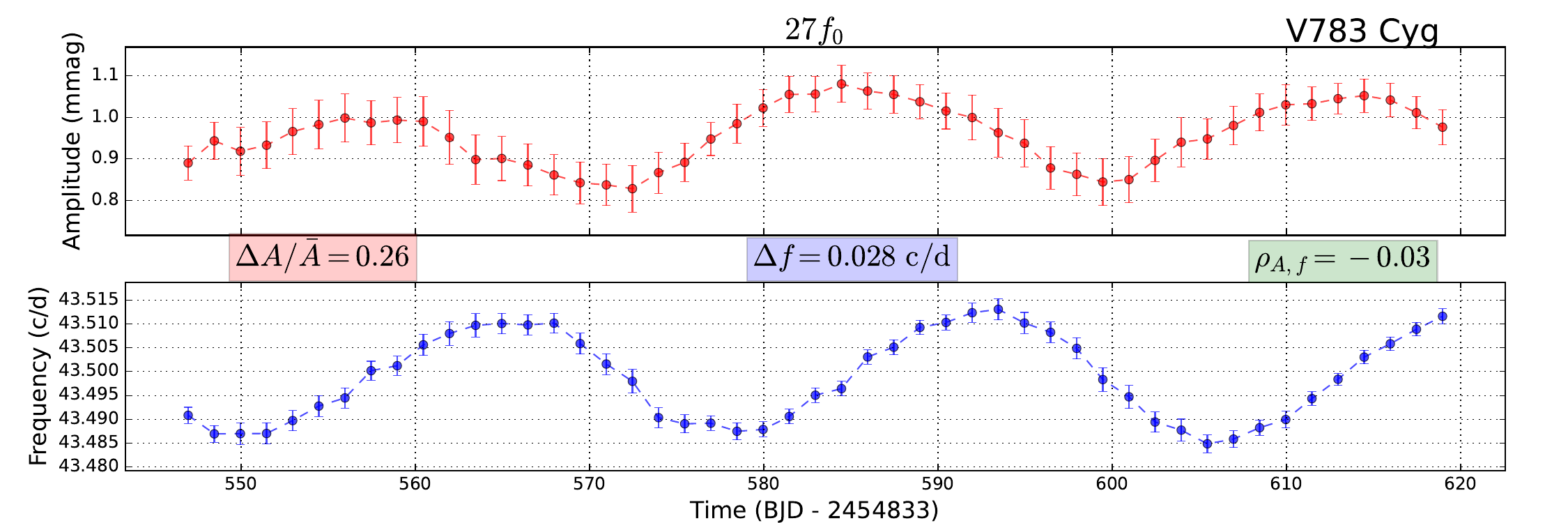}
  \includegraphics[width=0.48\textwidth]{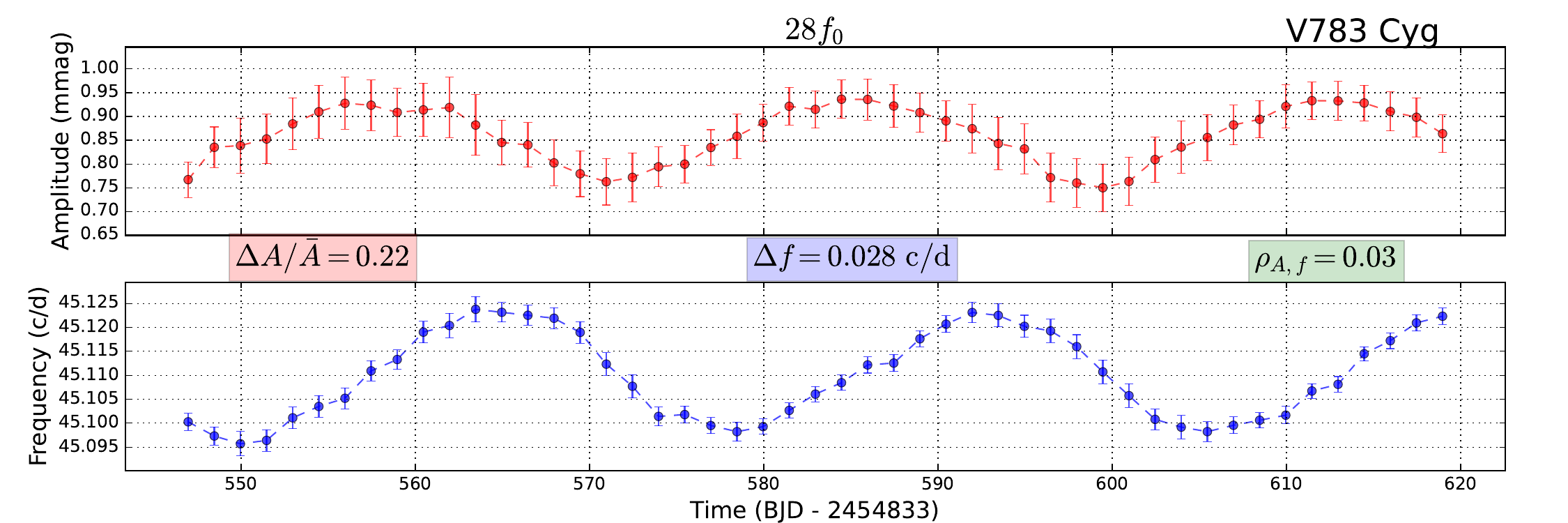}
  \caption{Temporal variations in amplitude and frequency for harmonics $f_0$--$47f_0$, part II.}
  \label{fig:var_amp_freq02}
\end{figure*}

\begin{figure*}[htp]
  \centering
  \includegraphics[width=0.48\textwidth]{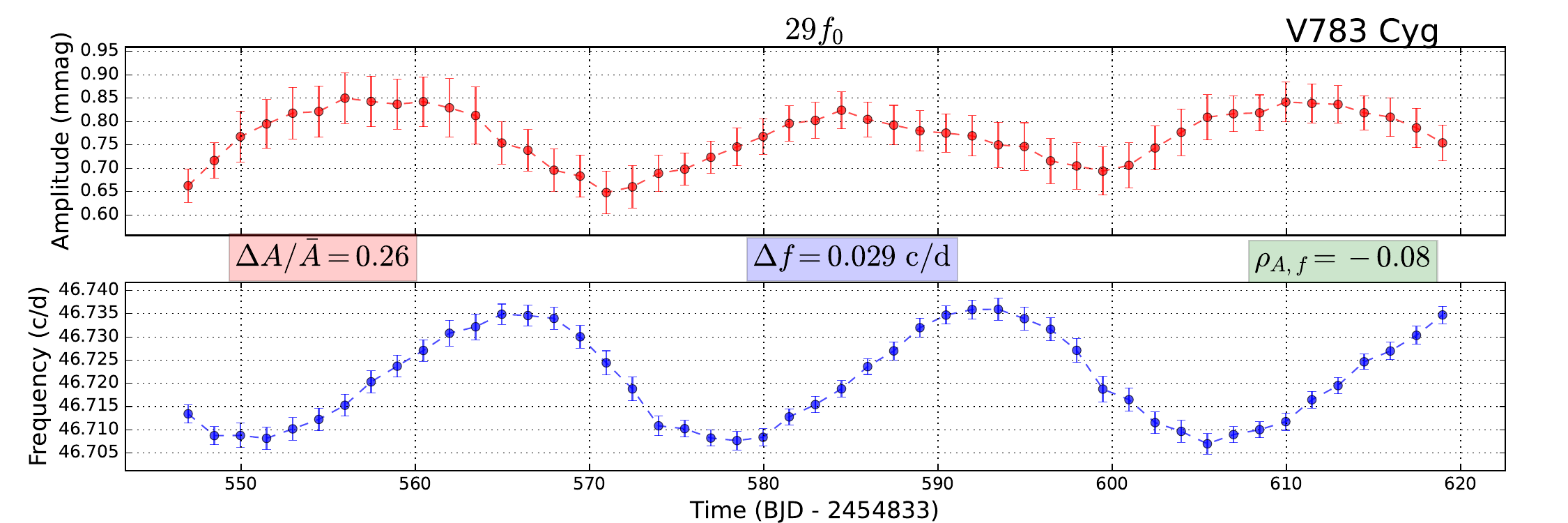}
  \includegraphics[width=0.48\textwidth]{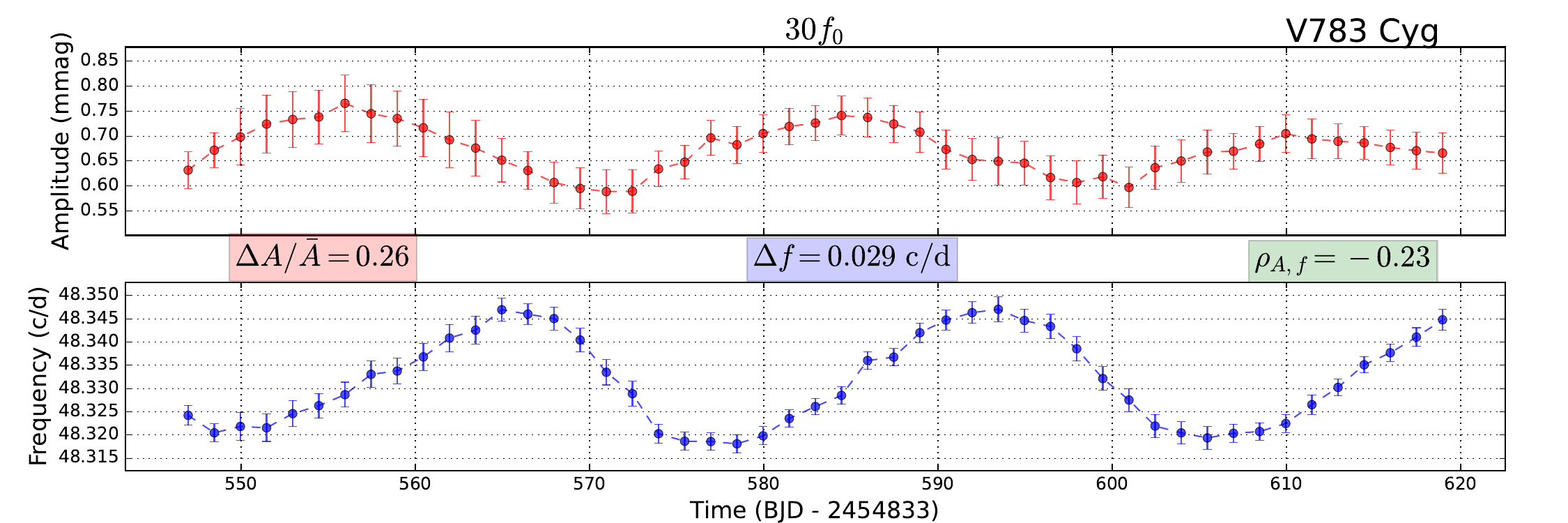}
  \includegraphics[width=0.48\textwidth]{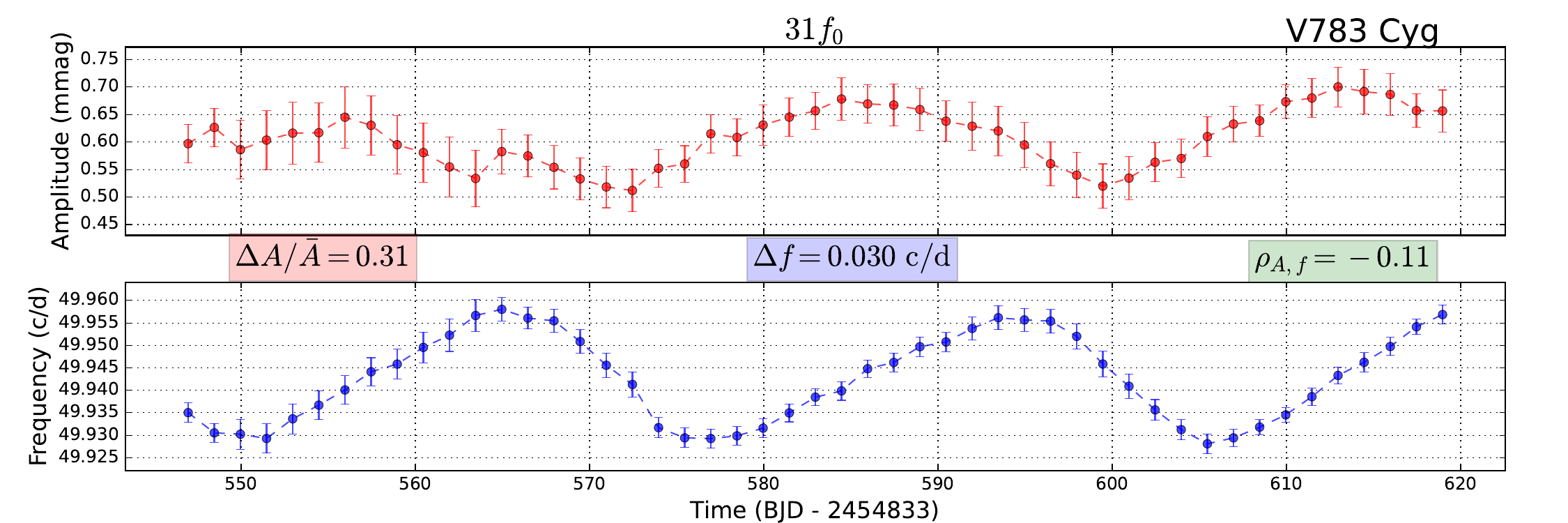}
  \includegraphics[width=0.48\textwidth]{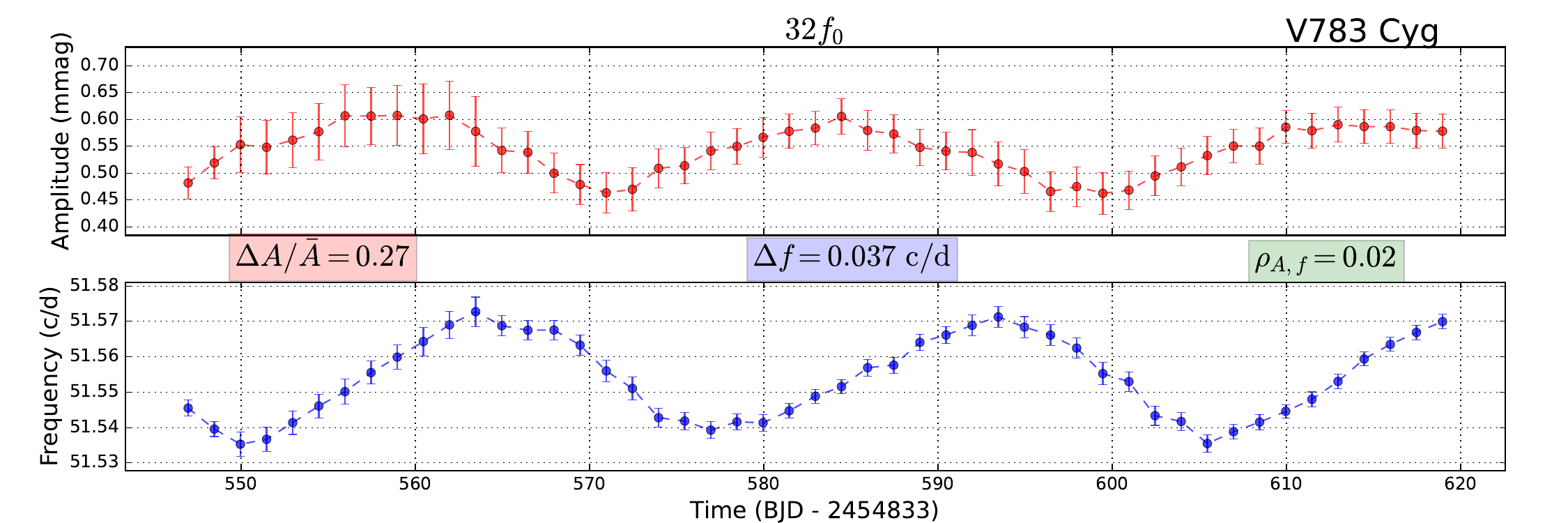}
  \includegraphics[width=0.48\textwidth]{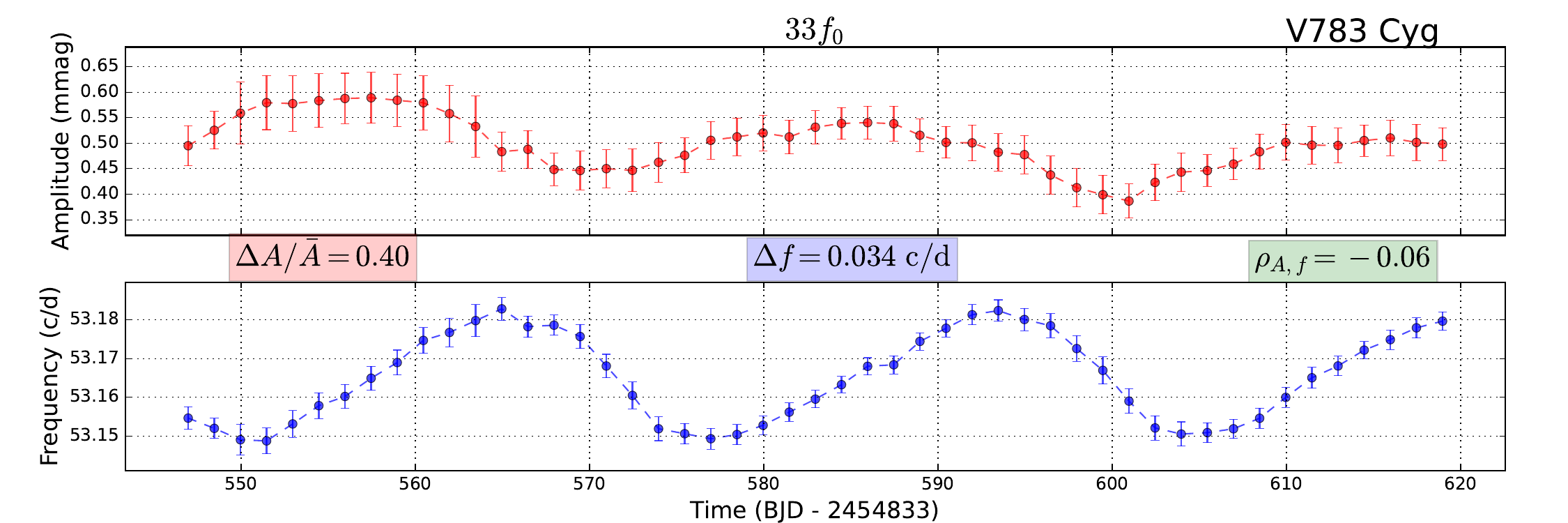}
  \includegraphics[width=0.48\textwidth]{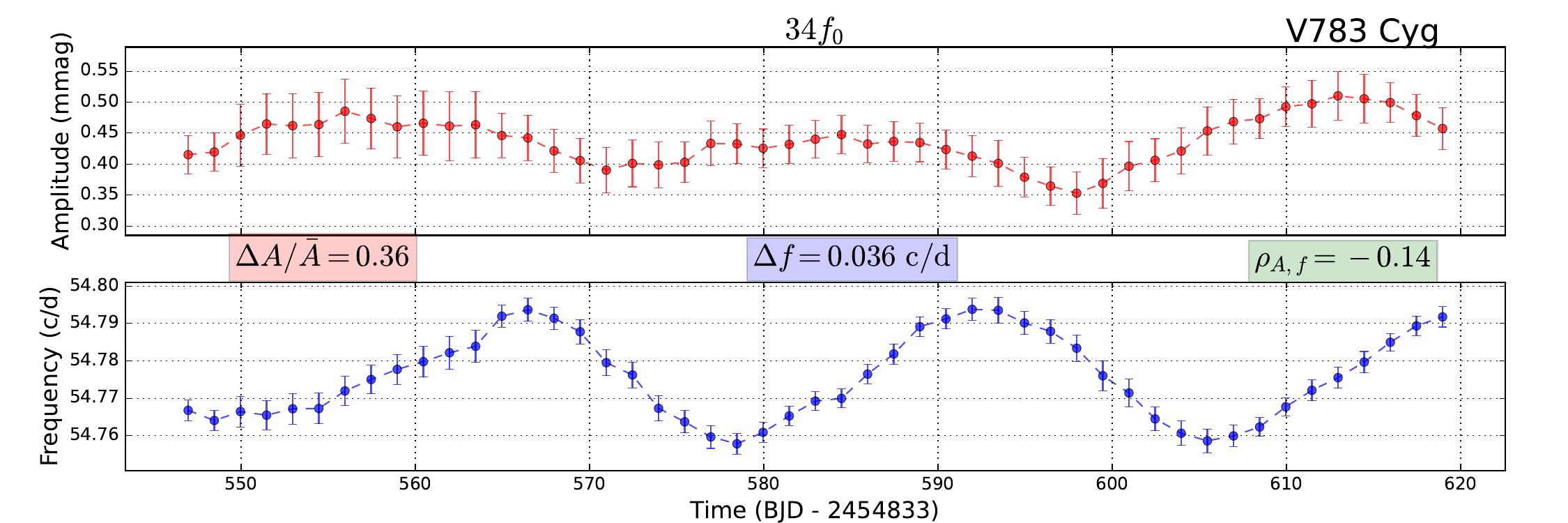}
  \includegraphics[width=0.48\textwidth]{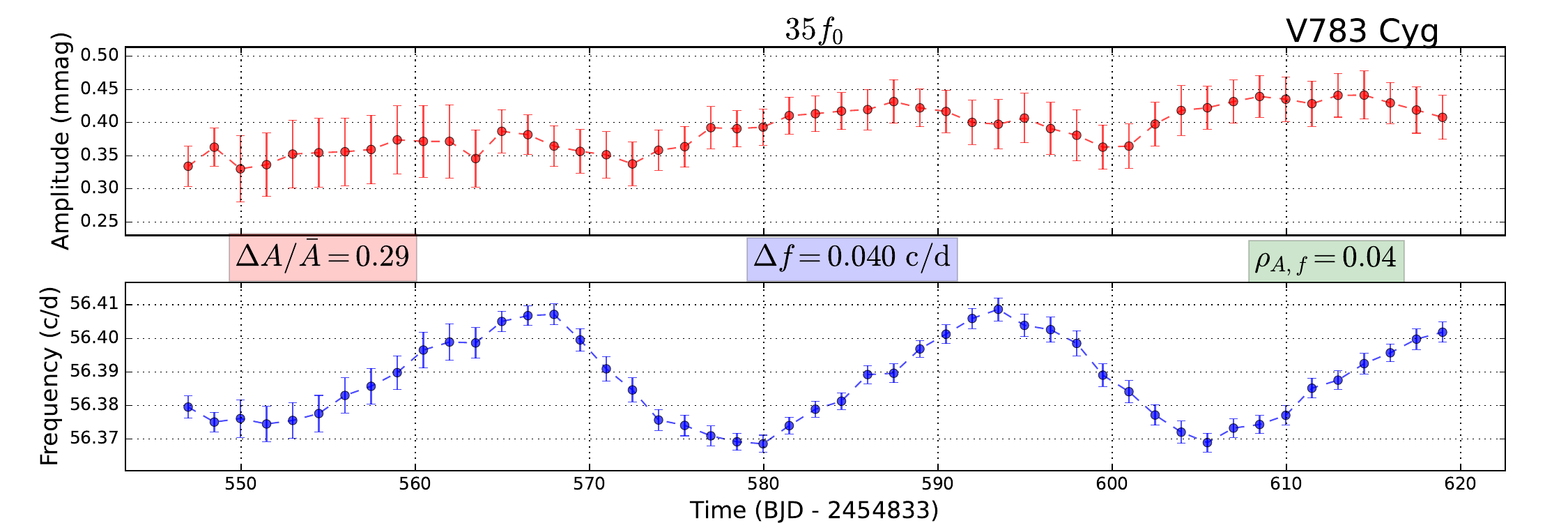}
  \includegraphics[width=0.48\textwidth]{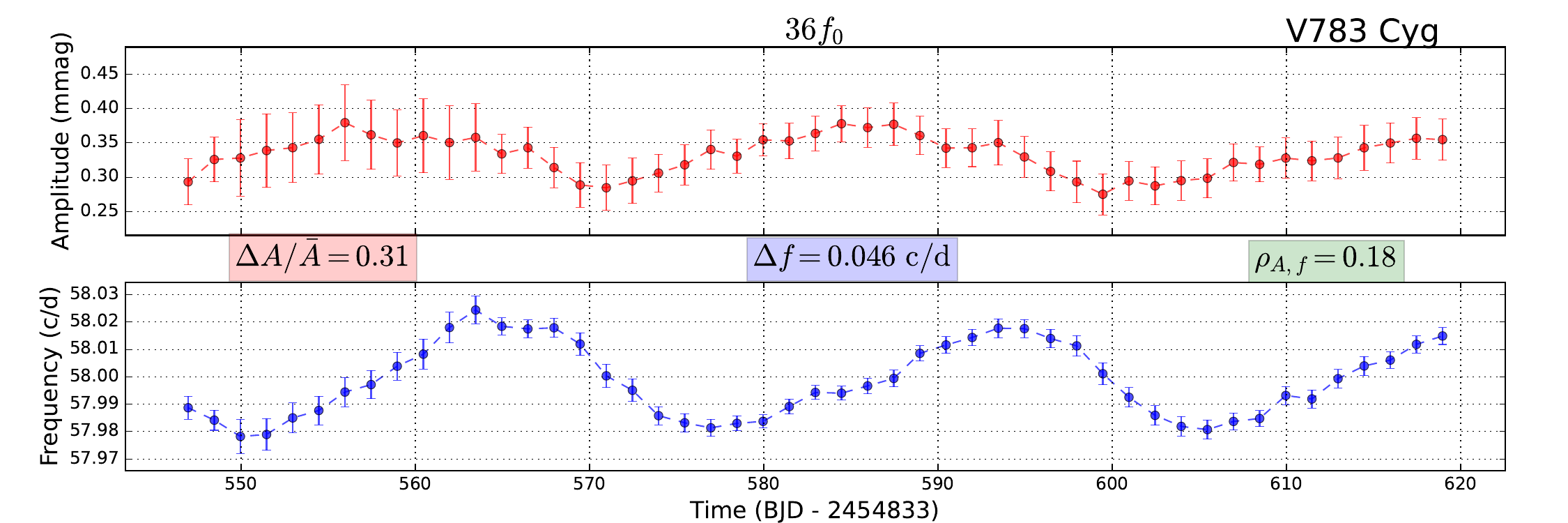}
  \includegraphics[width=0.48\textwidth]{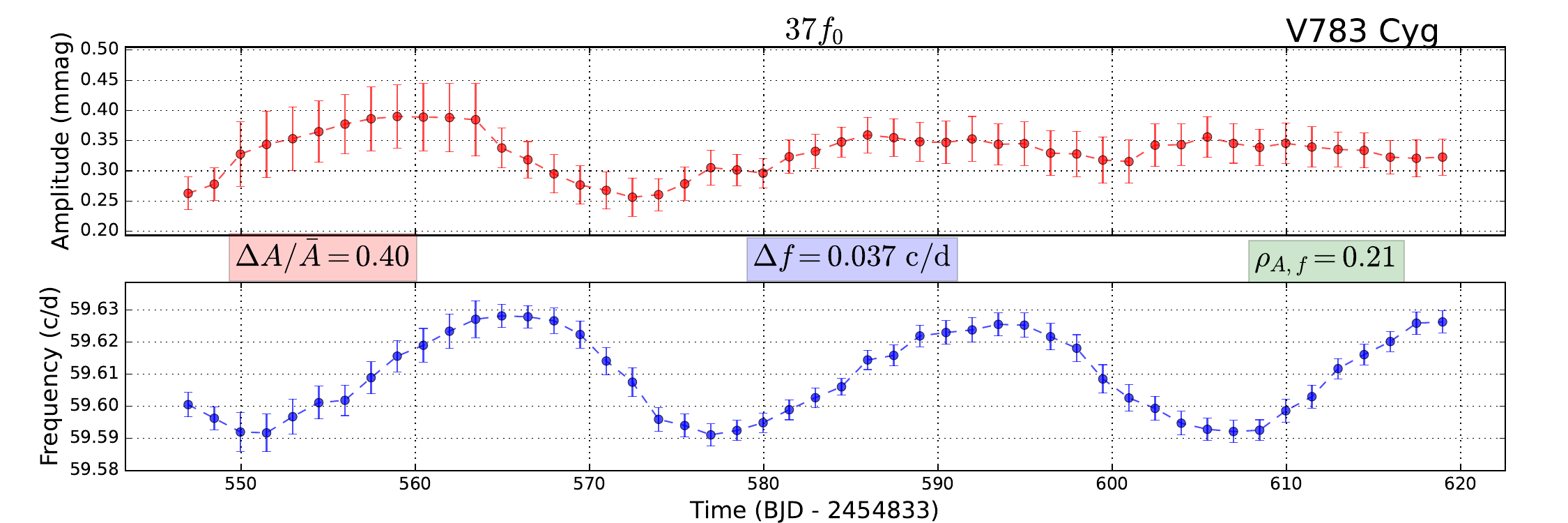}
  \includegraphics[width=0.48\textwidth]{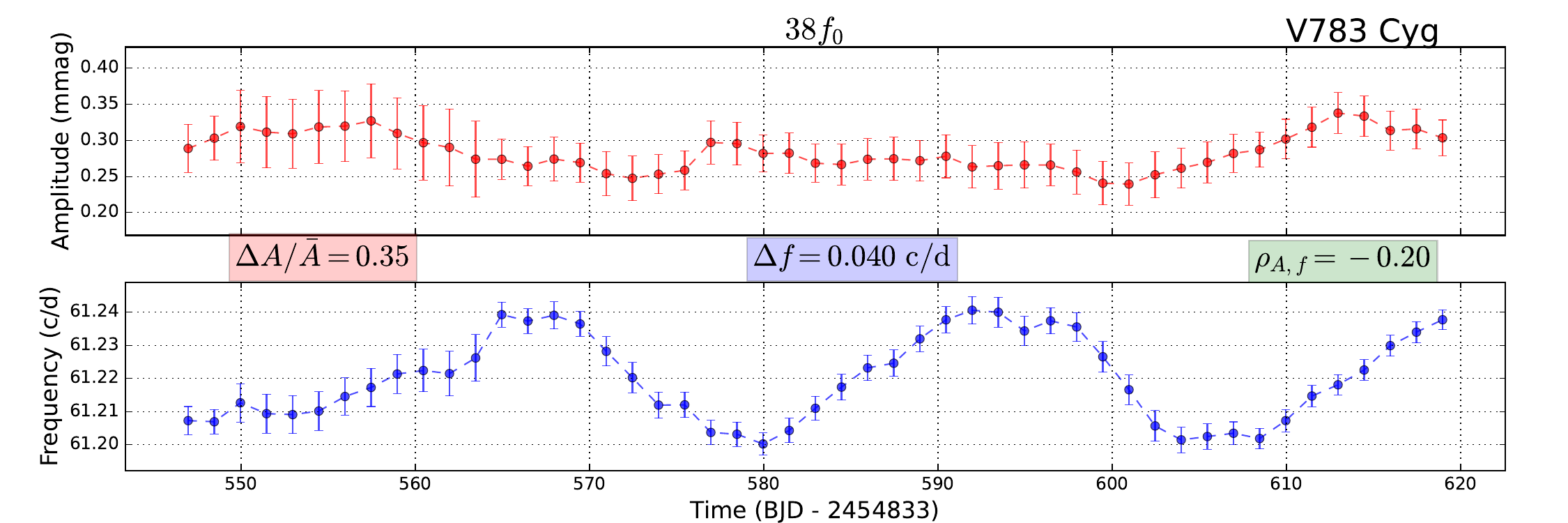}
  \includegraphics[width=0.48\textwidth]{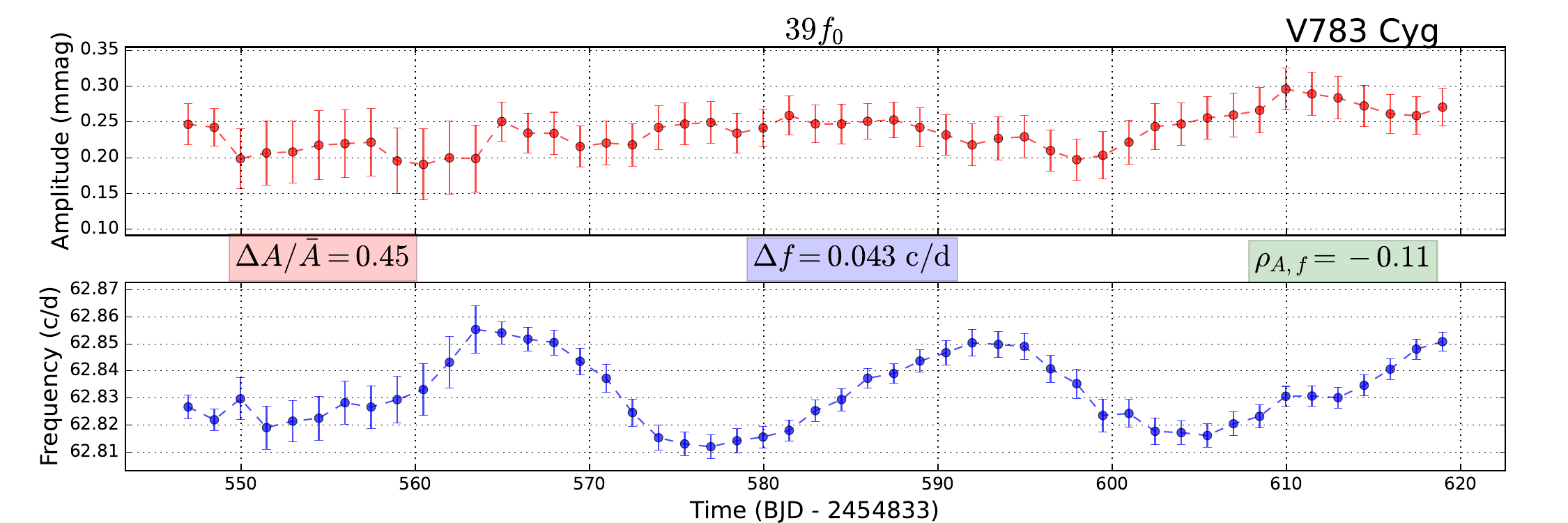}
  \includegraphics[width=0.48\textwidth]{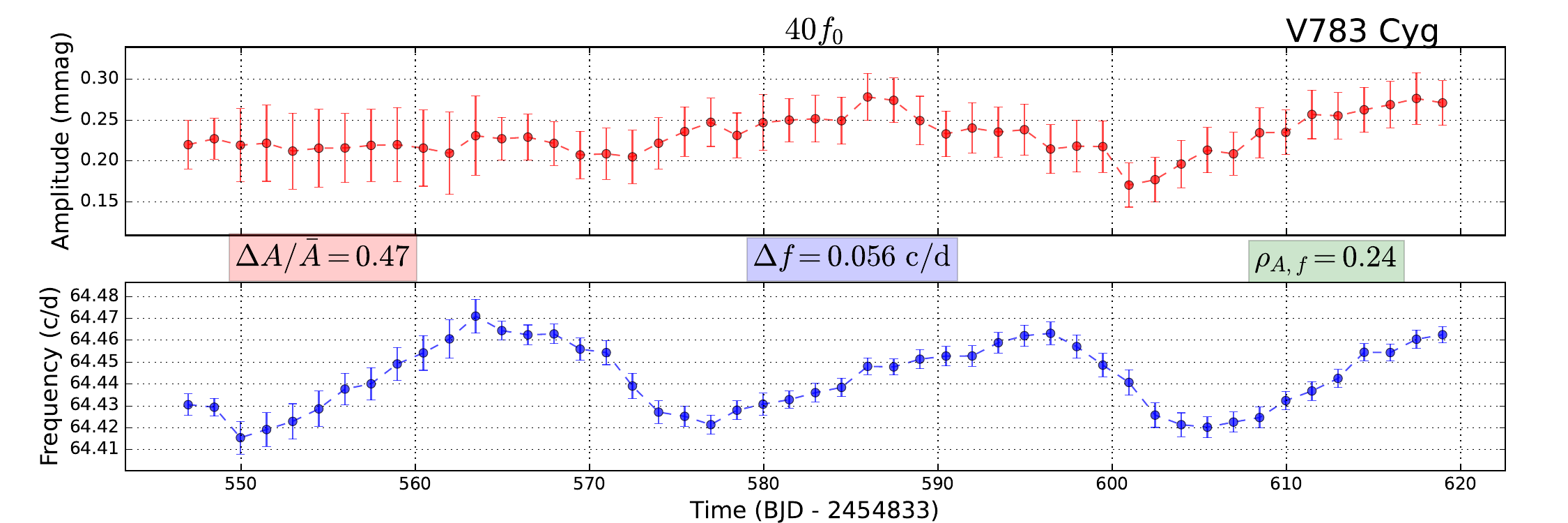}
  \includegraphics[width=0.48\textwidth]{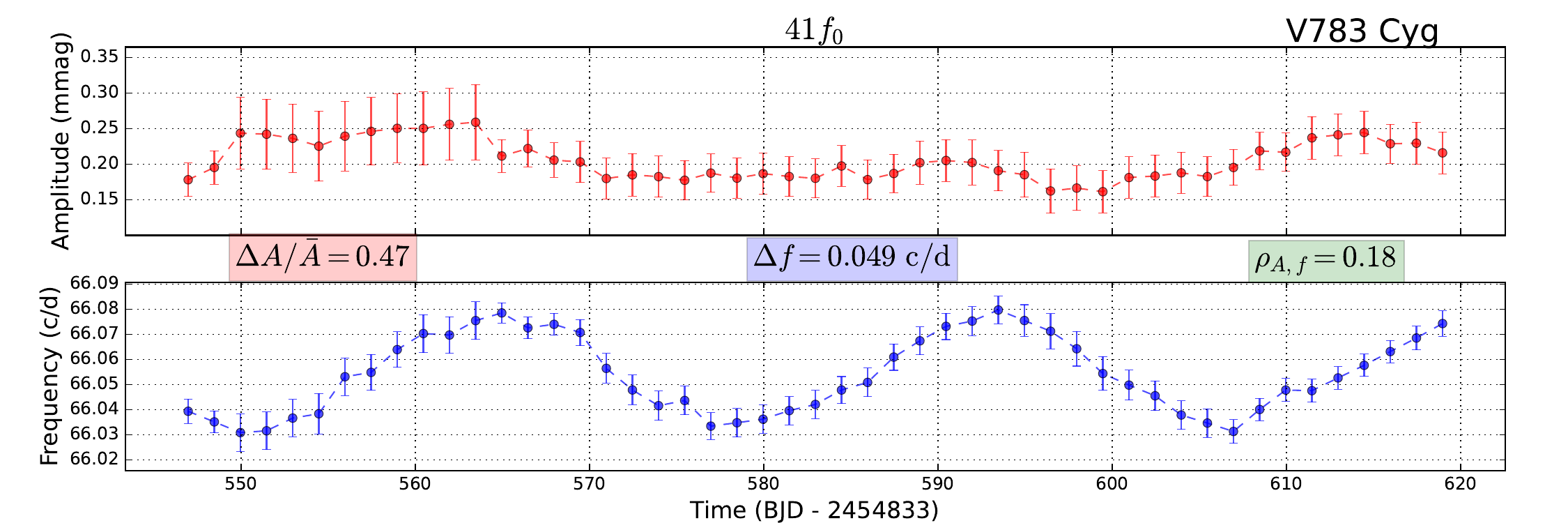}
  \includegraphics[width=0.48\textwidth]{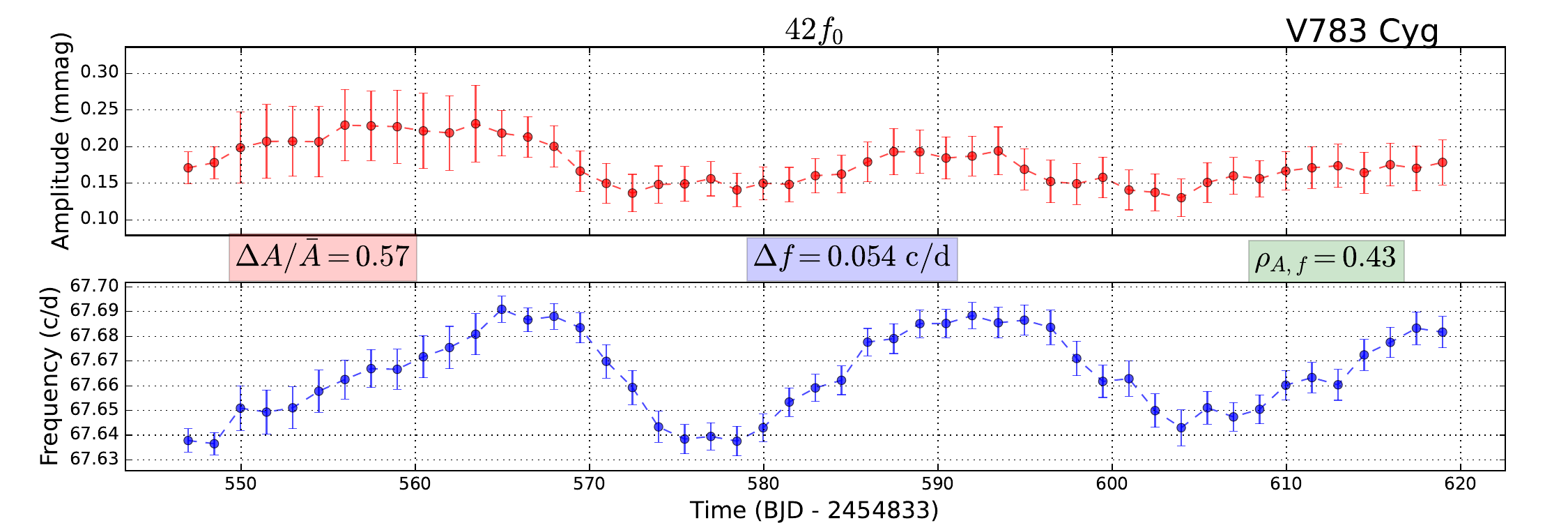}
  \caption{Temporal variations in amplitude and frequency for harmonics $f_0$--$47f_0$, part III.}
  \label{fig:var_amp_freq03}
\end{figure*}

Following the methodology established in our previous work \citep{Niu2023, Niu2024, Xue2024, Niu2025, Niu2026_RRab_non}, we extracted the amplitudes and frequencies of the harmonics $f_0$--$47f_0$ (Table~\ref{tab:harmonics}) and their temporal variations over $\sim$90 days (Figures~\ref{fig:var_amp_freq01}--\ref{fig:var_amp_freq03}).
To extract the amplitudes and frequencies of the harmonic pulsation modes, we performed a standard iterative pre-whitening procedure, continuing until the signal-to-noise ratio (S/N) fell below 7.0 (Table~\ref{tab:harmonics}).

To investigate temporal variations in the amplitudes and frequencies of the harmonics, we employed a short-time Fourier transform (STFT) approach \citep{Niu2023, Niu2024, Xue2024, Niu2025, Niu2026_RRab_non}.
A sliding window of 15 days was advanced across the full dataset ($\sim$90 days) in increments of 1.5 days (Figures~\ref{fig:var_amp_freq01}--\ref{fig:var_amp_freq03}).
The window length of 15 days represents a compromise between frequency resolution (needed to resolve individual harmonics at high orders) and temporal resolution (needed to track variations on the Blazhko timescale); it yields approximately 18 sampling points per Blazhko cycle.

\subsection{Quantifying Variability and Correlations}
\label{app:methods02}
To quantify the amplitude and frequency variations and their interrelations, we computed the following metrics:
\begin{itemize}
\item \textbf{Relative amplitude variation}: $\Delta A/\bar{A} \equiv (A_\mathrm{max} - A_\mathrm{min})/\bar{A}$, where $A_\mathrm{max}$ and $A_\mathrm{min}$ are the maximum and minimum amplitude values from the STFT, and $\bar{A}$ is the mean amplitude over the full dataset. This metric normalizes the amplitude range, allowing comparison across harmonics of vastly different mean amplitudes.
\item \textbf{Absolute frequency variation}: $\Delta f \equiv f_\mathrm{max} - f_\mathrm{min}$, where $f_\mathrm{max}$ and $f_\mathrm{min}$ are the maximum and minimum frequency values. This is reported in absolute terms (c/d) because the frequency variations are of similar magnitude across all harmonics.
\item \textbf{Pearson correlation coefficient} $\rho_{A,f}$: quantifies the linear correlation between the amplitude and frequency variations. A value of $+1$ indicates perfect correlation, $-1$ indicates perfect anti-correlation, and $0$ indicates no linear relationship.
\end{itemize}
These quantitative measures are displayed in Figures~\ref{fig:var_amp_freq01}--\ref{fig:var_amp_freq03}.

\clearpage
\section{Additional Tables}
\label{app:tables}
\startlongtable
\begin{deluxetable*}{l|ccccccc}
  \label{tab:harmonics}
  \tablecaption{Observational parameters of the harmonics detected in V783~Cyg.}
  \tablehead{\colhead{ID} & \colhead{Frequency} & \colhead{Frequency error} & \colhead{Amplitude} & \colhead{Amplitude error} & \colhead{S/N} \\
  \colhead{} & \colhead{(c/d)} & \colhead{(c/d)} & \colhead{(mmag)} & \colhead{(mmag)} & \colhead{}}
  \startdata
$f_0$	&  1.61112&	0.00004 &  274    & 2	 & 144.3 \\
$2f_0$	&  3.22220&	0.00004 &  146    & 1	 & 140.4 \\
$3f_0$	&  4.83328&	0.00004 &   93.0  & 0.6	 & 145.3 \\
$4f_0$	&  6.44437&	0.00004 &   56.9  &	0.4	 & 155.8 \\
$5f_0$	&  8.05545&	0.00004 &   27.9  &	0.2	 & 140.4 \\
$6f_0$	&  9.66659&	0.00004 &   17.1  &	0.1	 & 138.7 \\
$7f_0$	& 11.27767&	0.00004 &    9.95 &	0.07 & 137.4 \\
$8f_0$	& 12.88870&	0.00007 &    4.42 & 0.05 & 	89.7  \\
$9f_0$	& 14.49950&	0.00008 &    2.59 &	0.04 & 	73.0  \\
$10f_0$	& 16.11070&	0.00008 &    3.22 &	0.04 & 	81.7  \\
$11f_0$	& 17.72189&	0.00007 &    3.83 &	0.04 & 	89.7  \\ 
$12f_0$	& 19.33297&	0.00007 &    3.97 &	0.04 &  88.5  \\
$13f_0$	& 20.94411&	0.00008 &    4.01 &	0.05 & 	73.0  \\
$14f_0$	& 22.55525&	0.00006 &	 3.87 &	0.04 & 	97.1  \\
$15f_0$	& 24.16633&	0.00006 &    3.59 &	0.04 & 101.6 \\
$16f_0$	& 25.77742&	0.00006 &    3.30 &	0.03 &	99.7  \\
$17f_0$	& 27.38850&	0.00006 &    3.02 &	0.03 & 102.2 \\ 
$18f_0$	& 28.99958&	0.00006 &    2.72 &	0.03 &	97.6  \\
$19f_0$	& 30.61072&	0.00006 &    2.40 &	0.03 &	94.7  \\
$20f_0$	& 32.22180&	0.00007 &    2.15 &	0.02 &	90.0  \\
$21f_0$	& 33.83283&	0.00007 &    1.91 &	0.02 &	84.3  \\
$22f_0$	& 35.44391&	0.00007 &    1.69 &	0.02 &	83.1  \\
$23f_0$	& 37.05494&	0.00008 &    1.48 &	0.02 &	77.2  \\
$24f_0$	& 38.66619&	0.00009 & 	 1.31 &	0.02 &	70.6  \\
$25f_0$	& 40.27727&	0.00009 &    1.17 &	0.02 &	68.0  \\
$26f_0$	& 41.8882 &	0.0001  &    1.02 &	0.02 &	64.5  \\
$27f_0$	& 43.4994 &	0.0001  &    0.90 &	0.02 &	58.5  \\
$28f_0$	& 45.1105 &	0.0001  &    0.80 &	0.02 &	53.2  \\ 
$29f_0$	& 46.7216 &	0.0001  &    0.71 &	0.01 &	48.3  \\
$30f_0$	& 48.3327 &	0.0001  &    0.63 &	0.01 &	43.8  \\
$31f_0$	& 49.9435 &	0.0001  &    0.57 &	0.01 &	41.8  \\
$32f_0$	& 51.5550 &	0.0002	&    0.49 &	0.01 &	35.9  \\
$33f_0$	& 53.1659 &	0.0002	&    0.45 &	0.01 &	33.5  \\
$34f_0$	& 54.7764 &	0.0002	&    0.40 &	0.01 &	29.7  \\
$35f_0$	& 56.3880 &	0.0002	&    0.35 &	0.01 &	26.7  \\
$36f_0$	& 57.9996 &	0.0003	&    0.29 &	0.01 &	22.9  \\
$37f_0$	& 59.6102 &	0.0003	&    0.29 &	0.01 &	22.1  \\ 
$38f_0$	& 61.2210 &	0.0003	&    0.25 &	0.01 &  19.5  \\
$39f_0$	& 62.8323 &	0.0004	&    0.22 &	0.01 &  16.6  \\
$40f_0$	& 64.4438 &	0.0004	&    0.20 &	0.01 &	15.8  \\
$41f_0$	& 66.0548 &	0.0004	&    0.17 &	0.01 &	14.0  \\
$42f_0$	& 67.6656 &	0.0005	&    0.14 &	0.01 &	11.4  \\
$43f_0$	& 69.2768 &	0.0006	&    0.14 &	0.01 &	10.3  \\
$44f_0$ & 70.8885 &	0.0008	&    0.10 &	0.01 &	7.8   \\
$45f_0$	& 72.4993 &	0.0007	&    0.11 &	0.01 &	8.4   \\  
$46f_0$	& 74.1097 &	0.0008	&    0.10 &	0.01 &	7.7   \\
$47f_0$	& 75.7201 &	0.0008	&    0.10 &	0.01 &	7.5   \\
\enddata
\end{deluxetable*}

\clearpage
\section{Additional Figures}
\label{app:figures}

\begin{figure*}[htp]
  \centering
  \includegraphics[width=0.48\textwidth]{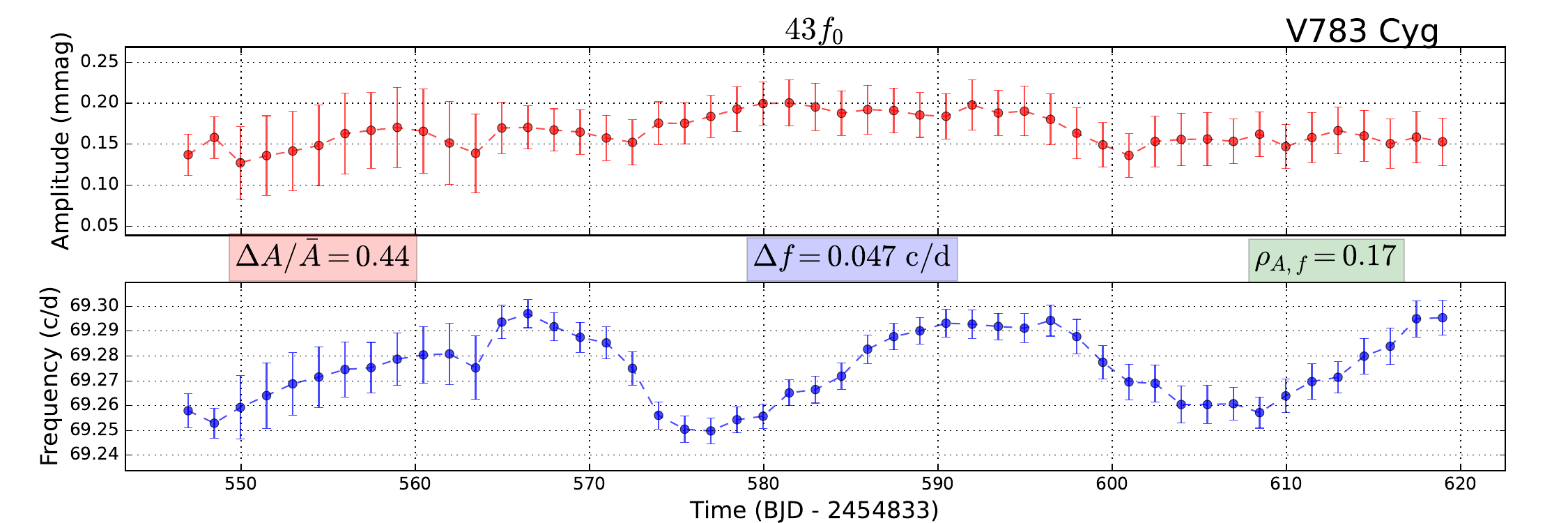}
  \includegraphics[width=0.48\textwidth]{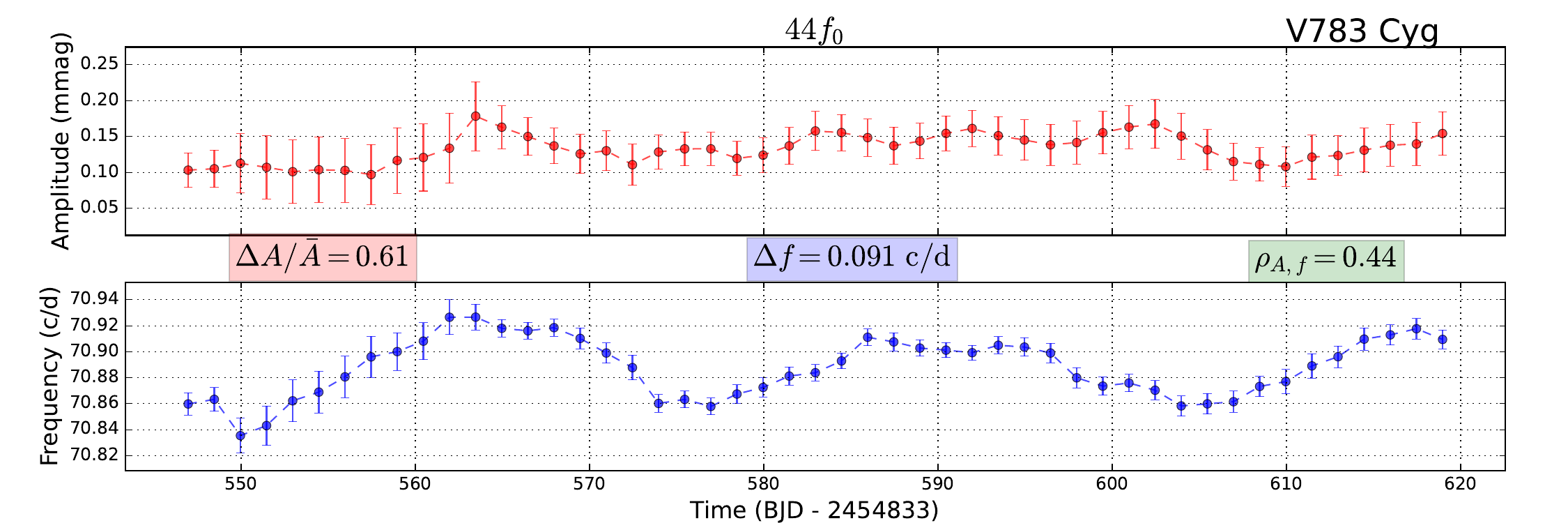}
  \includegraphics[width=0.48\textwidth]{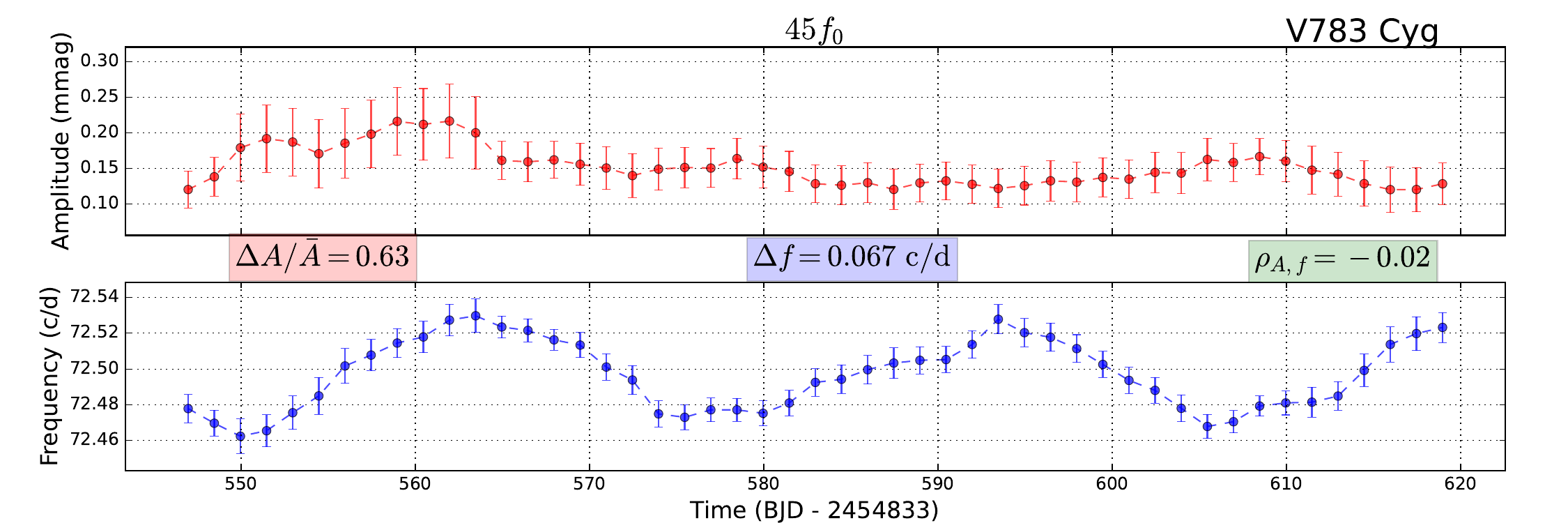}
  \includegraphics[width=0.48\textwidth]{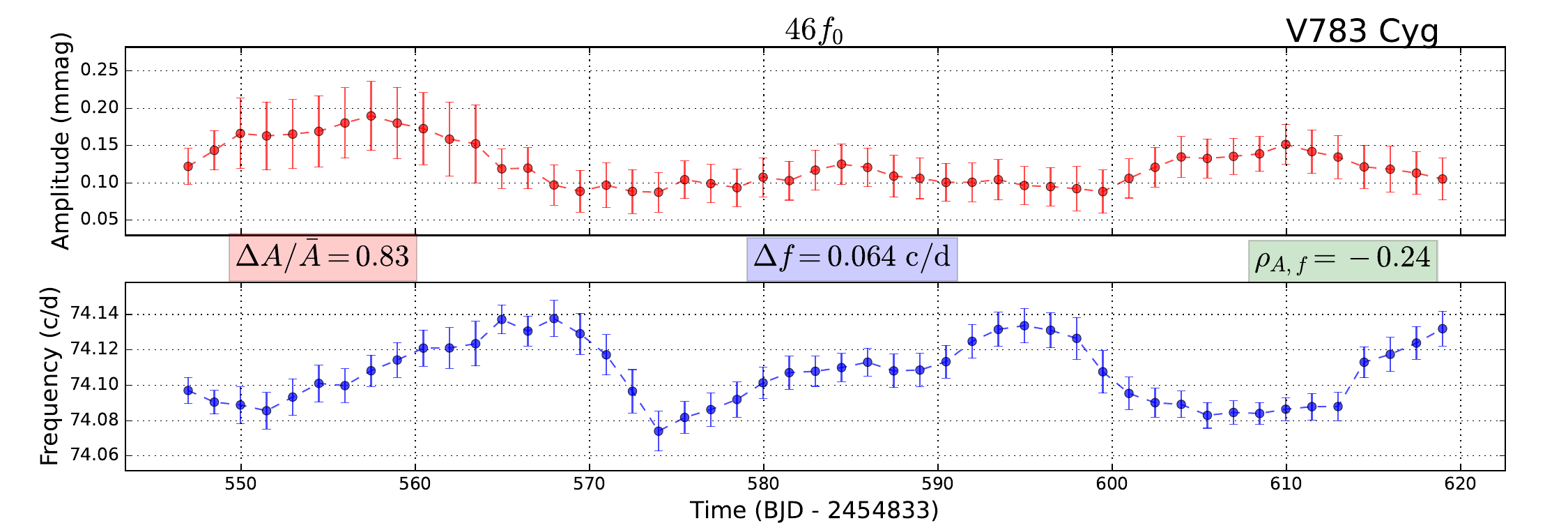}
  \includegraphics[width=0.48\textwidth]{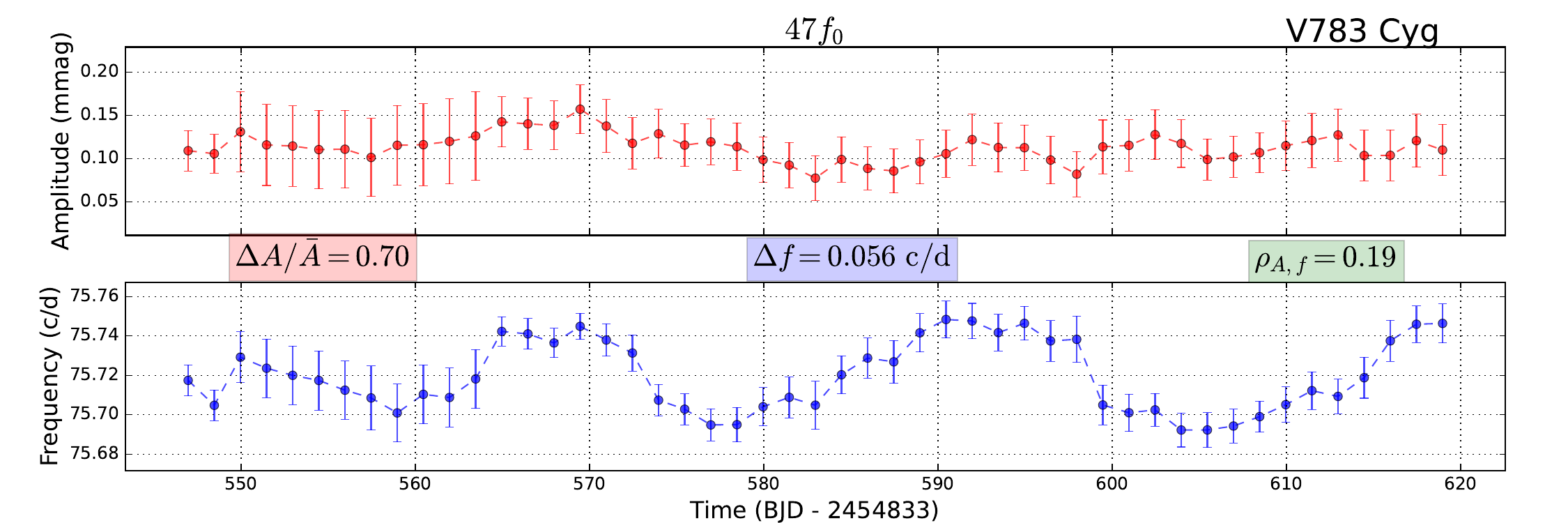}
  \caption{Temporal variations in amplitude and frequency for harmonics $f_0$--$47f_0$, part IV.}
  \label{fig:var_amp_freq04}
\end{figure*}

\begin{figure*}[htp]
  \centering
  \includegraphics[width=0.9\textwidth]{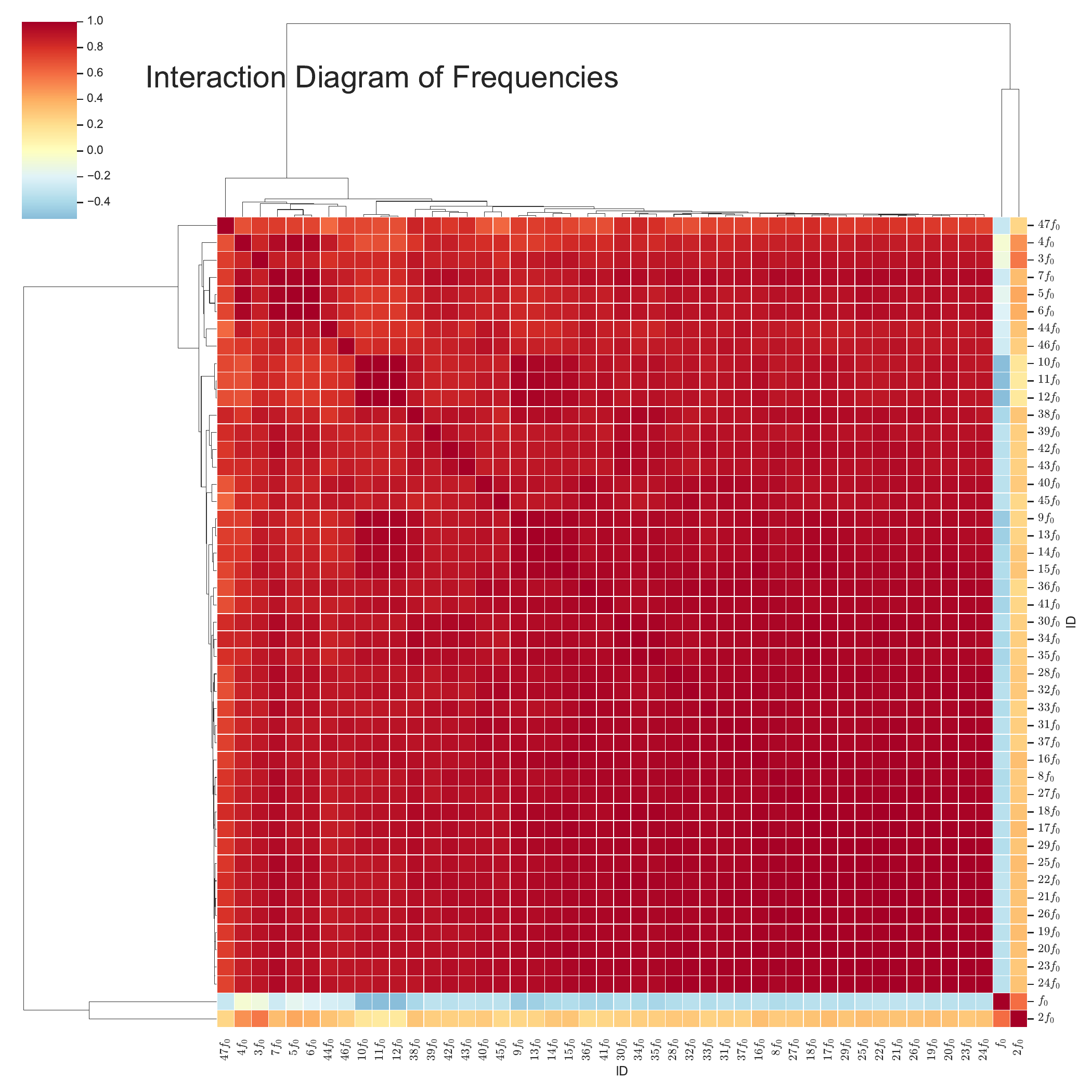}
  \caption{Interaction diagram of frequencies based on $f_0$ and its harmonics up to $47f_0$ in V783~Cyg.}
  \label{fig:ID_freq}
\end{figure*}

\end{CJK*}
\end{document}